\documentclass[%
 reprint,
 amsmath,amssymb,
 aps,
 showkeys
]{revtex4-2}

\usepackage{graphicx}
\usepackage{dcolumn}
\usepackage{bm}

\def\BeA{\begin{eqnarray}}
\def\EeA{\end{eqnarray}}

\begin{document}
	
	\preprint{APS/123-QED}
	
	\title{Exploring Weak measurements within the Einstein-Dirac Cosmological framework 
	} 
	
	\author{Williams Dhelonga-Biarufu}
	\email{williams.dhelonga@unamur.be}
	\author{Dominique Lambert}%
	\email{dominique.lambert@unamur.be}
	\affiliation{%
		Namur Insitute for Complex Systems (naXys),  Department of Mathematics, University of Namur, Rue de Bruxelles 61, B-5000 Namur, Belgium 
			}%

	\date{\today}
	
	\begin{abstract}
	
	Our study applies the Two-State Formalism alongside weak measurements within a spatially homogeneous and isotropic cosmological framework, wherein Dirac spinors are intricately coupled to classical gravity. We compute the weak values of the energy-momentum tensors, the Z component of spin, and the pure states. Weak measurements are a generalization and extension of the computation already made by Finster and Hainzl \cite{finster2011spatially} in A spatially homogeneous and isotropic Einstein-Dirac cosmology. Our analysis reveals that the acceleration of the Universe expansion can be understood as an outcome of post-selection, underscoring the effectiveness of weak measurement as a discerning approach for gauging cosmic acceleration.

	\end{abstract}

	
		\maketitle

	\section{\label{sec:level1}INTRODUCTION 
		}
Measuring involves assigning numerical values to the attributes or characteristics of a phenomenon. The impact of the observer on measurement results is a complex consideration in scientific research. In quantum mechanics, the alteration of a quantum state after its measurement is well known, but its interpretation is not obvious. For example, the famous double-slit experiment demonstrates that entities like electrons can behave as particles and waves, and measurement prescribes their behaviour \cite{kipnis1991history, aharonov2017finally}.

Weak measurements offer a procedure for measuring the system state through a weak coupling between the measurement apparatus and the system state, leaving the measured system almost undisturbed \cite{aharonov2005two}. The paid cost is that the pointer variable is not sharp but instead has a broad spread. Weak measurements have enabled, for instance, the characterization of a particle's average trajectory in the double-slit experiment without disrupting the interference pattern \cite{aharonov2017finally, tamir2013introduction, kocsis2011observing, rozema2012violation}. Unlike classical or ideal measurements, which result in the stochastic collapse of the system state into one of the eigenstates of the measured observable \cite{ferraz2022geometrical}, weak measurements do not induce a collapse of the state vector. Instead, they introduce a small bias to the state vector, and the measurement device exhibits a superposition of multiple values rather than a clear eigenvalue \cite{tamir2013introduction}. Consequently, weak measurements unveil unconventional weak values, including complex numbers.

In the literature, a few authors have explored weak measurements and the \emph{Two-state vector formalism} within the theoretical framework of cosmology and ontology. George Ellis and Rotman \cite{Elliscryst2010} proposed a paradigm that challenges conventional notions of time and reality, termed the
\emph{Crystallizing Block Universe (CBU)}, an extension of the \emph{Emergent Block Universe (EBU)}. According to their perspective, past, present, and future coalesce, amalgamating space-time into a singular entity. The future is conceptualized as a superposition of a myriad of possibilities, while the past remains immutable. The transition between these temporal states predominantly unfolds in the present, albeit in a non-uniform manner. Within the CBU framework, certain discrete patches of quantum indeterminacy endure and are resolved later on. Furthermore, CBU models integrate the \emph{Two-time interpretation} \cite{aharonov2005two} of quantum mechanics. Nevertheless, this theoretical and ontological paradigm lies beyond the purview of our current study.

Davies  \cite{davies2014quantum} was the first to propose applying weak measurements theory combined with pre- and post-selection to quantum cosmology and to explore the potential large-scale cosmological effects arising from this new sector of quantum mechanics. He illustrated the theory with a two-dimensional spacetime toy model of a scalar field with mass $m$ propagating in an expanding Universe with the scale factor $a(t)$ and metric $ ds^2 = dt^2 -a(t)dx^2 $. He resolved a debate regarding the coupling of pre- and post-selected quantum fields to gravity and proposed an experimental test for weak values in cosmology. He observed that, in the literature, the equation  $ G_{\mu \nu} + \text{higher order terms in curvature }=  \frac{\langle 0_{fin} |T_{\mu \nu} | 0_{in}\rangle  }{\langle 0_{fin}| 0_{in}\rangle }$ originally derived by DeWitt, who adapted the Schwinger effective action theory of quantum electrodynamics to the gravitational case, the source term is nothing else than the weak value of the stress-energy-momentum tensor $ T_{\mu \nu}$, \cite{dewitt1965dynamical}. This equation was widespread in 1970, \cite{boulware1975quantum, hartle1979quantum, cisowski2022colloquium}.

After Davies's paper  \cite{davies2014quantum}, no research papers have been pursued in the direction of the theory of weak measurements in Cosmology until now. The subject seems to arouse interest. Charis Anastopoulos has just published on ar$\chi$iv a paper on \emph{Final States in Quantum Cosmology: Cosmic Acceleration as a Quantum Post-selection Effect}, \cite{anastopoulos2024final} wherein he argues that there is no compelling physical reason to preclude a probability assignment with a final quantum state at the cosmological level and analyses its implications in quantum cosmology. One significant result is that cosmic acceleration emerges as a quantum post-selection effect. 

We aim to examine weak measurements of Dirac particles within the framework of time symmetry as applied to the Einstein-Dirac system in a homogeneous and isotropic space such as the Friedman-Lemaître-Robertson-Walker(FLRW) space. To our knowledge, no prior research has explored this direction. Drawing upon insights from the paper of Finster and Hainzl, titled \emph{A Spatially Homogeneous and Isotropic Einstein-Dirac Cosmology} \citep{finster2011spatially}, we endeavor to extend specific findings to the domain of weak measurements. Among the outcomes of our investigation are the computations of weak values, including those about energy-momentum, pure states, and the Z component of spin. Additionally, we demonstrate that cosmic acceleration in the Universe may be regarded as a consequence of post-selection. This corroborates earlier studies that have explored the potential of spinor fields to elucidate phenomena such as the inflationary period in the early Universe and, subsequently, the concept of dark energy. Notably, Anastopoulos arrived at a similar conclusion without invoking a spinor field or dark energy, thus emphasizing alternative avenues for understanding the observed cosmological acceleration. Ribas reached the same conclusion without resorting to weak measurements or post-selection techniques. References to relevant literature supporting these assertions include \cite{ribas2005fermions, ochs1993fermions, saha2005spinor, ribas2007cosmological, rakhi2010cosmological,kremer2013cosmological}.

In the time-symmetric formulation of quantum mechanics \cite{aharonov1964time}, a quantum system is characterized by two state vectors at a specific time instead of a single one. The pre-selected state is determined by measurements conducted on the system prior to $t_0$, while the post-selected state is by measurements at times $t > t_0$. The post-selected state is conceptualized as a quantum state evolving backward in time. Weak measurements occur in the interval between these conventional measurements through a weak coupling between the measurement apparatus and the system \cite{tamir2013introduction}.

Formally, if $\hat{A}$ is a Hermitian operator on the system $S$, $|\psi_{in}\rangle$ and $|\psi_{out}\rangle$ (or $|\psi_{fin}\rangle$) are the pre-selected and post-selected state vectors in the Hilbert space of $S$, and $|\phi(q)\rangle$ is the state vector of the needle of the measurement device, $ g(t)$ a weak coupling impulse such that $ \int_{0}^{T} g(t)dt=g_0T$, $T$ the coupling time and $\hat{P}_d$ the momentum operator conjugated to the position operator $\hat{Q}_d$,  then weakly measuring an ensemble of pre-selected vectors states $|\psi_{in}\rangle$ with the interaction Hamiltonian  $\hat{H}=g(t)\hat{A} \otimes \hat{P}_d$ will yield
\begin{align*}
|\psi_{w}\rangle &= e^{(-i\hat{H} T)}|\psi_{in}\rangle \otimes |\phi(x)\rangle\\
&=\sum_i \alpha_i |a_i\rangle \otimes |\phi(q-g_0Ta_i)\rangle.
\end{align*}
wherein$|\psi_{in}\rangle =\sum_i \alpha_i |a_i\rangle$ is expressed in vector basis of the observable $\hat{A}$.  Making strong measurement on the device pointer state will yield $\langle q |\psi_{w}\rangle= \sum_i \alpha_i \phi(q-g_0Ta_i) |a_i\rangle $. 

Weak measurement with post-selection is obtained by making a projection with the post-selected state $\langle \psi_{fin}|$. In detail, let $\hat{P} = |\psi_{fin}\rangle \langle \psi_{fin}|\otimes I$ be an operator which projects onto the post-selected state $ |\psi_{fin}\rangle $, and  $|\psi_{w}\rangle = e^{(-i\hat{H} T)}|\psi_{in}\rangle \otimes |\phi(x)\rangle$ the entangled state of the system with the state of the measurement device \cite{tamir2013introduction},\cite{aharonov2008two}, 
 then we get
\begin{align*}
	\hat{P} | \psi_{w}\rangle &=|\psi_{fin}\rangle \langle \psi_{fin}|e^{(-i\hat{H} T)}|\psi_{in}\rangle \otimes |\phi(x)\rangle.
\end{align*}
Hence, weak coupling between the system and the device will yield: 
\begin{align*}
	&\hat{P} | \psi_{w}\rangle\approx |\psi_{fin}\rangle \langle \psi_{fin}|(1 - i \hat{A} \otimes \hat{P}_d \cdot g_0T)|\psi_{in}\rangle \otimes |\phi(q)\rangle\\
	&=|\psi_{fin}\rangle  \langle \psi_{fin}|\psi_{in}\rangle e^{-i \left(\langle \hat{A} \rangle_w \hat{P}_d T \right)}\otimes |\phi(q)\rangle
\end{align*}
where $ \langle \hat{A} \rangle_w = \frac{\langle \psi_{fin} |\hat{A} |\psi_{in}\rangle  }{\langle \psi_{fin}|\psi_{in}\rangle } $  is the so-called weak value of the observable $ \hat{A}$.
The probabilities to get this result is $| \langle \psi_{fin}|\psi_{in}\rangle| ^2 $.

Weak values are weak measurement results and must be understood as statistical averages. They are also members of a decomposition of eigenvalues weighted by the relative amplitudes of finding the out states among the ensemble of states vectors identically prepared in $|\psi_{in}\rangle$. A weak value should then be understood as a conditional probability where one of the events undergoes a perturbation. Dressel \cite{dressel2014colloquium} clarifies this probabilistic understanding of weak values. He shows that weak values characterize the relative correction to a detection probability $|\langle \psi_{fin}|\psi_{in} \rangle|^2 $ due to a small intermediate perturbation $\hat{U}(\epsilon)$ that results in a modified detection probability $|\langle \psi_{fin}|\hat{U}(\epsilon)|\psi_{in} \rangle|^2 $. Alternatively, in a simple way, he defines weak values as complex numbers that one can assign to the powers of a quantum observable operator $\hat{A} $ using two states: an initial state $|\psi_{in}\rangle$ and a final state $|\psi_{fin}\rangle$. Weak values might lie outside the spectrum of eigenvalues and can be complex numbers. One interesting phenomenon is the amplification obtained when the pre-selected state and the post-selected one are almost orthogonal \cite{tamir2013introduction, davies2014quantum}. 

In cosmology \cite{davies2014quantum}, all observations and measurements are considered weak in the quantum sense due to the nature of the processes involved. For instance, when observing the redshift of a galaxy, the measurement is conducted by observing the light emitted from a large number of photons originating from numerous sources within the galaxy. While the emission of a single photon from an atom may not be considered weak from the atom's perspective, the use of a large ensemble of photons to measure a property of the entire galaxy constitutes a weak measurement in the quantum sense. This is because the quantum back-reaction of the photons on the relevant physical variable of the entire galaxy, such as its momentum, is negligible. Therefore, the large-scale nature of cosmological observations and measurements, involving statistical averages over a large ensemble of photons, results in weak values in the quantum sense.  

This idea of weakness of all observations and measurements in cosmology is rooted in classical field. Dressel, in his paper \citep{dressel2014classical} have shown that, given two spacetime hypersufaces $ \sigma_I$ and $\sigma_F $, on which field states $| I\rangle $ and $\langle F|$ are define as boundary conditions, the classical background field $\phi$ has the form of a weak value of the quantum field operator $\phi=\frac{\langle F | \hat\phi | I \rangle}{\langle F | I \rangle}$. Any weak interaction, that is any weak measurement  that does not appreciably perturb the classical background field or its boundary conditions will result in a quantum weak measurement whether or not the measurement coupling is weak for every field excitation.

 In the laboratory, the Two-State vectors require human interventions. In quantum cosmology, there are many proposals, as noticed by Davies: Hartle and Hawking came up with the no-boundary wave function, which is the state of the Universe before the Planck epoch \cite{Hartle1983WaveFO}. In a semi-classical approach, in the framework of the theory of quantum fields,  the initial state is taken to be a vacuum state. As for the final state, this can be anything. In weak measurement, one will ensure that the final state is not orthogonal to the in-state.

 For the Einstein-Dirac in FLRW context, the pre-selected state will be the spinor solution of the Einstein-Dirac equation in the limit of the massless Universe, that is, the radiation Universe, whereas the final state will be the spinor solution of the Einstein-Dirac equation for the dust Universe. The implementation of a weak measurement in this specific context is intended to acquire geometrical information about the Universe. The subsequent computational analysis is made feasible by the association between weak measurement and the Berry phase  \cite{ferraz2022geometrical}. This has never been explored before, as far as we know.

Regarding the Einstein-Dirac system, it investigates the interaction between particles with a spin of $1/2$ and the gravitational field. The Einstein field equations essentially relate the geometry of spacetime to the distribution of matter and energy within that spacetime. In simpler terms, they describe how matter and energy, represented here by the stress tensor applied on spinors, influence the curvature of spacetime and how the curvature of spacetime influences the motion of matter and energy. Finster \cite{finster2011spatially} investigated the nonlinear coupling of gravity to matter in a time-dependent, spherically symmetric Einstein-Dirac system. He showed that quantum oscillations of the Dirac wave functions can prevent the formation of a big bang or big crunch singularity.

Weak measurements present numerous advantages, foremost among them being the capability to detect exceedingly subtle effects while minimizing disturbance to the system state. Onur Hosten and Paul Kwiat \cite{hosten2008observation} exemplified the utility of weak measurement techniques in amplifying the spin Hall effect of light, Meng-Jun Hu  \cite{hu2017gravitational} discussed in his paper the proposal for Weak Measurements Amplification based Laser Interferometer Gravitational-wave Observatory (WMA-LIGO) to detect gravitational waves using weak measurements to amplify ultra-small phase signals,  while Dixon et al. \cite{dixon2009ultrasensitive} applied these methods to enhance the detection of minute transverse deflections in an optical beam. This facet of weak measurements proves highly advantageous in cosmology, enabling the amplification and measurement of wave functions originating in the remote past, close to singularities.
		
This paper is organized as follows: after the introduction, we give a summary of the Einstein-Dirac system and then find the spinor solutions for the radiation Universe and the dust one. In section III, we give the results of our main computations on weak measurements on the Einstein-Dirac solutions. In section IV, we give a conclusion. The appendices provide more details of our computations.
%

	\section{The Einstein-Dirac Solution in Friedmann-Lemaître-Robertson-Walker Space}
	
	In this section, we provide an overview of the Einstein-Dirac solution within the Friedmann-Lemaître-Robertson-Walker (FLRW) space framework. For a more comprehensive treatment, readers are encouraged to refer to \cite{finster2011spatially} and \cite{finster2009dirac}.
	
The Einstein-Dirac (ED) equations are given by
	\begin{equation}
		R^{i}_{j} - \frac{1}{2}R\delta^{i}_{j} = 8\pi\kappa T^{i}_{j}, \quad (\mathcal{D}-m)\Psi=0, \label{eq:ed-equation}
	\end{equation}
where $\kappa$ represents the gravitational constant, $R^{i}_{j}$  the Ricci tensor, $R$ the scalar curvature,  $T^{i}_{j}$  the energy-momentum tensor of the Dirac particles, $\{i, j \}$  the indices representing the coordinates $\{t, r, \theta, \phi \}$, $\mathcal{D}$  the Dirac operator, $m$  the mass of the Dirac particles, and $\Psi$  the Dirac wave function.
	
	Our Einstein-Dirac equations are formulated within the FLRW space, a spacetime manifold with a (1,3) signature. This simple cosmological model is characterized by a metric described by the homogeneous and isotropic line element
	\begin{equation}
		ds^{2} = dt^{2} - R^{2}(t) \left(\frac{dr^{2}}{1 - kr^{2}} + r^{2}d\Omega^{2}\right), \label{eq:flrw-metric}
	\end{equation}
	where $t \in \mathbb{R}$ represents time, $(\theta, \phi) \in (0, \pi) \times [0, 2\pi)$ are angular coordinates, $r$ is the radial coordinate, $R(t)$ the scale function, $d\Omega^{2}$ the line element on $S^{2}$, and $k$ can take values of -1, 0, or 1, corresponding to open, flat, and closed universes, respectively. We adopt "Planck units" where $c=\hbar=G=1$.
	
	In this article, we specifically consider the case of a closed universe with $k=1$, although similar computations can be extended to other cases. In this scenario, the line element simplifies to
	\begin{equation}
		ds^{2} = dt^{2} - R^{2}(t) \left(\frac{dr^{2}}{1 - r^{2}} + r^{2}d\Omega^{2}\right). \label{eq:closed-metric}
	\end{equation}
	
	In this case, the coordinate $r$ will vary in the interval $ (0,1)$. After conducting detailed calculations, outlined in the appendices of \cite{finster2011spatially} and \cite{finster2009dirac}, the Dirac operator takes the following form
	\begin{equation}
		\left[i\gamma\left(\partial_{t} + \frac{3}{2}\frac{\dot{R}}{R}\right) - R(t)m + \begin{pmatrix}
			0 & \mathcal{D}_{\mathcal{H}} \\
			-\mathcal{D}_{\mathcal{H}} & 0
		\end{pmatrix}\right]\Psi = 0, \label{eq:dirac-operator}
	\end{equation}
	where $\mathcal{D}_{\mathcal{H}}$ represents the purely spatial operator on $S^{3}$, given by
	\begin{equation}
		\mathcal{D}_{\mathcal{H}} = i\sigma^r \left(\partial_r+ \frac{f - 1}{r f}\right) + i\sigma^{\theta}\partial_{\theta} + \sigma^{\phi}\partial_{\phi}, \label{eq:spatial-operator}
	\end{equation}
	where $ \sigma^r, \sigma^{\theta } \text{and} \sigma^{\phi} $ are linear combinations of the Pauli matrices defined as follows
	
\begin{equation}
	\left\{
	\begin{array}{lll}
		\sigma^r:= f(r)\left(\cos{\theta}\,\sigma^3 + \sin{\theta }\cos{\phi}\,\sigma^1 + \sin{\theta }\sin{\phi }\,\sigma^2\right)\\
		\sigma^{\theta}:= \frac{1}{r}\left(-\sin{\theta }\,\sigma^3 + \cos{\theta}\cos{\phi }\,\sigma^1 + \cos{\theta }\sin{\phi }\,\sigma^2\right)\\
		\sigma^{\phi}:=\frac{1}{r\sin{\theta}}\left( -\sin{\phi }\,\sigma^1 + \cos{\phi }\,\sigma^2\right).
	\end{array}
	\right.
	\label{eq: sigmas}
\end{equation}	
wherein the Pauli matrices are
\begin{equation}
\sigma^1 = \begin{pmatrix} 0 & 1 \\ 1 & 0 \end{pmatrix}, \quad
\sigma^2 = \begin{pmatrix} 0 & -i \\ i & 0 \end{pmatrix}, \quad
\sigma^3 = \begin{pmatrix} 1 & 0 \\ 0 & -1 \end{pmatrix}
\end{equation}\label{paulim}
	and the function $ f(r)$ is
	\begin{eqnarray}
	\begin{array}{lll}
		f(r) &=& \sqrt{1-r^2}, \hspace{1cm}  r \in (0,1).\\
	\end{array}
	\label{eq: function_f}
\end{eqnarray}
	This spatial operator $\mathcal{D}_{\mathcal{H}}$  has a purely discrete spectrum
	\begin{equation}
		\sigma(\mathcal{D}_{\mathcal{H}}) = \left\{ \pm \frac{3}{2}, \pm \frac{5}{2}, \pm \frac{7}{2}, \ldots \right\}, \label{eq:discrete-spectrum}
	\end{equation}
	and the dimension of the corresponding eigenspace is given as:
	\begin{equation}
		\text{dim}\; \text{ker}\left(\mathcal{D}_{\mathcal{H}} - \lambda\right) = \lambda^{2} - \frac{1}{4}. \label{eq:eigenspace-dimension}
	\end{equation}
	An orthonormal eigenvector basis in terms of spherical harmonics and Jacobi polynomials is provided in the appendix of \cite{finster2009dirac} and is denoted by $\psi^{\pm}_{njk}$, where $n \in \mathbb{N}_{0}$, $j \in \mathbb{N}_{0} + \frac{1}{2}$, and $k \in \{-j, -j + 1, \ldots, j\}$.
	
	The spatial differential operator on the eigenvector can be represented as follows
	\begin{equation}
		\mathcal{D}_{\mathcal{H}} \psi^{\pm}_{njk} = \lambda \psi^{\pm}_{njk}, \quad \text{where} \quad \lambda = \pm(n+j+1). \label{eq:eigenvector-operator}
	\end{equation}
	
	We represent the normalized eigenfunction of $\mathcal{D}_{\mathcal{H}}$ related to the eigenvalue $\lambda$ by $\psi_{\lambda} \in L^{2}(S^{3})^{2}$.
	
	To further analyze the system, we employ a separation ansatz for 
	\begin{equation}
		\Psi = \frac{1}{R(t)^\frac{3}{2}}\begin{pmatrix}
			\alpha(t)\psi_{\lambda}(r, \theta, \phi) \\
			\beta(t)\psi_{\lambda}(r, \theta, \phi)
		\end{pmatrix}. \label{eq:separation-ansatz}
	\end{equation}
	This ansatz enables the derivation of a coupled system of ordinary differential equations (ODEs) for the complex-valued functions $\alpha(t)$ and $\beta(t)$:
	\begin{equation}
		i\frac{d}{dt}\begin{pmatrix}
			\alpha(t) \\
			\beta(t)
		\end{pmatrix} = \begin{pmatrix}
			m & -\frac{\lambda}{R(t)} \\
			-\frac{\lambda}{R(t)} & -m
		\end{pmatrix} \begin{pmatrix}
			\alpha(t) \\
			\beta(t)
		\end{pmatrix}. \label{eq:coupled-odes}
	\end{equation}
	The spinors are normalized according to:
	\begin{equation}
		|\alpha|^2 + |\beta|^2 = \lambda^{2} - \frac{1}{4}. \label{eq:normalization}
	\end{equation}
	These spinors enter the Einstein equation via the energy-momentum tensor of the wave function, ensuring a coupling with the Dirac equation. The non-vanishing components of the energy-momentum tensor are computed in Appendix B of \cite{finster2011spatially} and are given by
	\begin{align}
		T^{0}_{0} &= R^{-3} \left[m\left(|\alpha|^2 - |\beta|^2\right) - \frac{2\lambda}{R} \text{Re}\left(\alpha\bar{\beta}\right)\right], \label{eq:energy-momentum-t00} \\
		T^{r}_{r} &= T^{\theta}_{\theta} = T^{\phi}_{\phi} = R^{-3} \frac{2\lambda}{3R} \text{Re}\left(\alpha\bar{\beta}\right). \label{eq:energy-momentum-trr}
	\end{align}
	
	A short calculation for the Einstein tensor $G^{j}_{k}$, given in \cite{finster2011spatially}, yields
	\begin{align}
		G_0^0 &= 3\left(\frac{\dot{R}^{2}+1}{R^{2}}\right), \label{eq:einstein-tensor-g00} \\
		G_r^r &= G_{\theta}^{\theta} = G_{\phi}^{\phi} = 2\frac{\ddot{R}}{R} + \frac{\dot{R}^{2}+1}{R^{2}}. \label{eq:einstein-tensor-grr}
	\end{align}

Knowing that :
	\begin{equation*}
		G^{i}_{j} = 8\pi\kappa T^{i}_{j},  \quad \text{where} \quad G^{i}_{j}=R^{i}_{j} - \frac{1}{2}R\delta^{i}_{j},
	\end{equation*}	
we set $\kappa=\frac{3}{8\pi}$ for a convenient computation as in \citep{finster2011spatially}. This yields the following expression, the Einstein equation
	\begin{equation}
		\dot{R}^2 + 1 = R^2T^{0}_{0},  \label{eq:einstein-dirac-equationb}
	\end{equation}
	
	which in terms of two level system is given as
	\begin{equation}
		\dot{R}^2 + 1 = \frac{m}{R} \left(|\alpha|^2 - |\beta|^2\right) - \frac{2\lambda}{R^2} \text{Re}\left(\alpha\bar{\beta}\right). \label{eq:einstein-dirac-equation}
	\end{equation}
	
	We derive also the acceleration of the Universe represented here by the second derivative with respect to time of the scale function given  firstly in terms of the energy-momentum tensors as
	
	\begin{equation}
		\ddot{R}  = \frac{R}{2} \left(3 T^{j}_{j} - T^{0}_{0}\right).  \label{eq:einstein-dirac-equationa2}
	\end{equation}	
	
	Then in terms of spinors we shall have
	
	\begin{equation}
		\ddot{R}  = -\frac{m}{2 R^2} \left(|\alpha|^2 - |\beta|^2\right) + \frac{2\lambda}{R^3} \text{Re}\left(\alpha\bar{\beta}\right). \label{eq:einstein-dirac-equation2}
	\end{equation}
	
	One sees immediately that the acceleration of the Universe is caused by the spinor field, and this is consistent along the lines of \cite{ribas2005fermions, ochs1993fermions, saha2005spinor, ribas2007cosmological, rakhi2010cosmological,kremer2013cosmological}. We will later on show some behaviours of the acceleration of the scale function.
	
	At this stage, we are left with two differential equations, equations (\ref{eq:coupled-odes}) and (\ref{eq:einstein-dirac-equation}), in which the spinor $\begin{pmatrix} \alpha \\ \beta \end{pmatrix}$ is considered as a two-level quantum state.
	
	As suggested by Finster and Hainzl in \cite{finster2011spatially}, the Einstein-Dirac equation can be rewritten in terms of a  \emph{Bloch vector} $\vec{v}$, where
	\begin{align}
		\vec{v} &= \begin{pmatrix}
			\langle \xi | \sigma^1 | \xi \rangle \\
			\langle \xi | \sigma^2 | \xi \rangle \\
			\langle \xi | \sigma^3 | \xi \rangle
		\end{pmatrix}, \label{eq:bloch-vector} \\
		\vec{b} &= 2\frac{\lambda}{R}\vec{e}_{1} - 2m\vec{e}_{3}. \label{eq:bloch-vector-b}
	\end{align}
	Here, $\xi$ represents the spinor $\begin{pmatrix} \alpha \\ \beta \end{pmatrix}$, $\sigma^{i}$ the Pauli matrices, and $\vec{e}_{i}$ the standard basis vectors in $\mathbb{R}^3$ ($i \in \{1, 2, 3\}$).
	
	The Einstein-Dirac equations can then be expressed as
	\begin{equation}
		\dot{\vec{v}} = \vec{b} \wedge \vec{v}, \quad \dot{R}^2 + 1 = -\frac{1}{2R} \vec{b} \cdot \vec{v}. \label{eq:einstein-dirac-bloch}
	\end{equation}
	Here, `$\wedge$' and `$\cdot$' denote the cross and scalar product in Euclidean space $\mathbb{R}^3$, respectively. We rewrite also the scale acceleration in terms of the Bloch vector components as
	\begin{equation}
		\ddot{R}  = -\frac{m}{2 R^2} v^3+ \frac{\lambda}{ R^3} v^1. \label{eq:einstein-dirac-equation3}
	\end{equation}
	
	To simplify the analysis, a rotation``U'' around the $\vec{e}_{2}$-axis is applied to $\vec{b}$, making $\vec{b}$ parallel to $\vec{e}_{1}$. This results in a transformed vector $\vec{w}$
	\begin{equation}
		\vec{w} = U\vec{v} = \begin{pmatrix}
			w^1 \\
			w^2 \\
			w^3
		\end{pmatrix} \label{eq:transformed-vector-w}
	\end{equation}
	where the components of $\vec{w}$ are given by
	\begin{small}
	\begin{align}
		w^1 &= \frac{1}{\sqrt{\lambda^2 + m^2 R^2}} \left(2\lambda \text{Re}\left(\alpha\bar{\beta}\right) - mR\left(|\alpha|^2 - |\beta|^2\right)\right), \label{eq:component-w1} \\
		w^2 &= 2\text{Im}\left(\alpha\bar{\beta}\right), \label{eq:component-w2} \\
		w^3 &= \frac{1}{\sqrt{\lambda^2 + m^2 R^2}} \left(2m \text{Re}\left(\alpha\bar{\beta}\right) + \lambda\left(|\alpha|^2 - |\beta|^2\right)\right). \label{eq:component-w3}
	\end{align}
	\end{small}
	
	At this stage, the Einstein-Dirac equations reduce to a system of ODEs involving the scale function $R(t)$ and the complex functions $\alpha(t)$ and $\beta(t)$
	\begin{equation}
		\dot{\vec{w}} = \vec{d} \wedge \vec{w}, \quad \dot{R}^2 + 1= -\frac{1}{R^2} \sqrt{\lambda^2 + m^2 R^2}	w^1, \label{eq:ode-system}
	\end{equation}
	where $\vec{d}$ is defined as
	\begin{equation}
		\vec{d} := \frac{2}{R} \sqrt{\lambda^2 + m^2 R^2}\vec{e}_{1} - \frac{\lambda mR}{\lambda^2 + m^2 R^2}\frac{\dot{R}}{R}\vec{e}_{2}. \label{eq:vector-d}
	\end{equation}
	
	The length of the Bloch vector is constant, and the normalization convention is
	\begin{equation}
		|\vec{w}| = \lambda^{2} - \frac{1}{4} = N. \label{eq:bloch-vector-length}
	\end{equation}
	
	Within weak measurement and the \emph{Two Vectors State Formalism}, this research primarily explores solutions to the Einstein-Dirac equation under specific conditions. These conditions involve either the mass parameter ($m=0$), in which case the solution of (\ref{eq:ode-system}) is the preselected state, or the eigenvalue ($\lambda=0$), in which case the solution of the same equation is the post-selected state.	
	
	In the case where $m=0$, the Einstein-Dirac equation simplifies to the well-known FLRW equation, describing the Universe's behaviour in the radiation-dominated phase. Conversely, in the second scenario,   $\lambda=0$ corresponds to a dust-dominated Universe in the second scenario. Notably, for sufficiently large-scale factor $R$, the Universe exhibits classical behaviour similar to the dust-dominated case.
	
	The above assumption can be linked easily from the complete and classical Friedmann general equation with $k=1, 0, -1$ corresponding to closed, flat and opened universe respectively
	\begin{equation}
	\dot{R}^2 +k=-\left(\frac{\rho_{m}}{R}+\frac{\rho_{r}}{R^2}-\Lambda R^2\right), \label{friedman-eq}
	\end{equation}
	where   $\rho_{m}$ is the density of matter, $\rho_{r}$ is the density of radiation and $\Lambda$ is the vacuum energy. For $R\ll 1$, in the right side of the expression (\ref{friedman-eq}), the radiative term dominates in the evolution of the Universe. The expression becomes 
	\begin{equation}
	\dot{R}^2 +k=-\left(\frac{\rho_{r}}{R^2}\right). \label{eq: friedman-eq1}
	\end{equation}
While, ignoring the vacuum energy, if $R\gg 1$, the matter term dominates and the expression (\ref{friedman-eq}) becomes
	\begin{equation}
	\dot{R}^2 +k=-\left(\frac{\rho_{m}}{R}\right), \label{eq: friedman-eq2}
	\end{equation}
	This equation describes a dust-dominated Universe.
	
	\begin{figure}[h]
		\includegraphics[scale=0.65, angle=0]{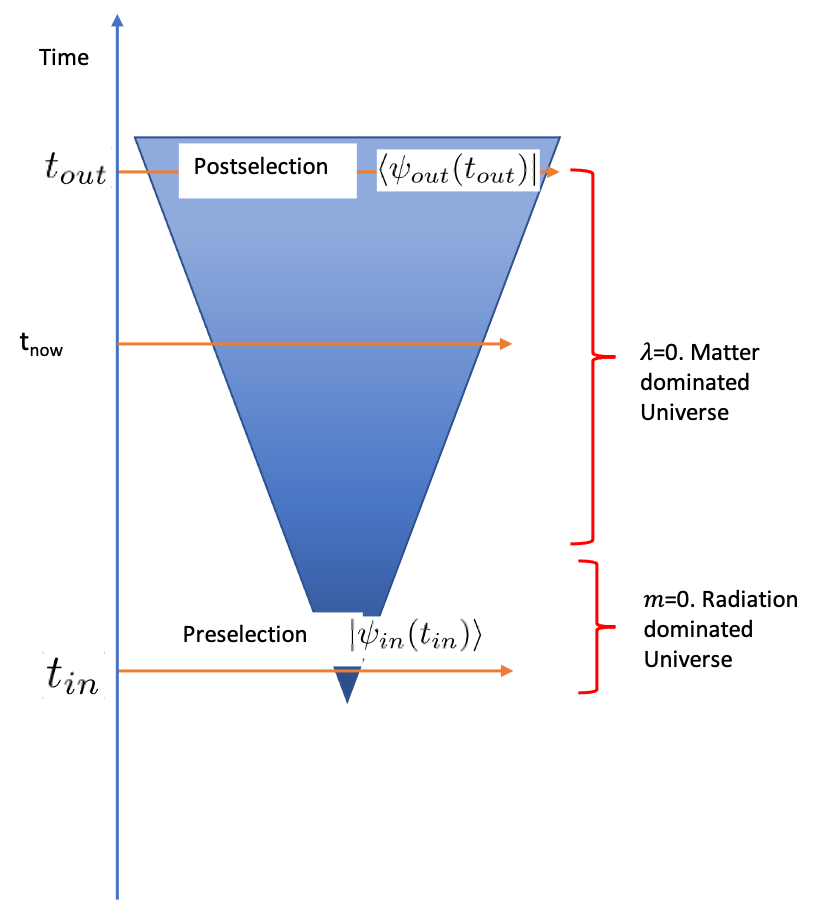}
		\centering
		\caption{pre-selection and post-selection à $t_{in}$ and $t_{out}$ }
	\end{figure}

	It is essential to note that the radiation-dominated Universe pertains to a brief period near the initial singularity. In contrast, the dust-dominated universe extends over a much longer timeframe, including the present day.

\subsection{$\mid \Psi_{in} \rangle$: Solution of the Einstein-Dirac equation for  $m=0$. }

When $m=0 $,  the ODEs in (\ref{eq:ode-system}) becomes

\begin{eqnarray}
	\left\{
	\begin{array}{lll}
		\dot{w}^1 &=& 0\\
		\dot{w}^2 &=& -\frac{2\lambda}{R}w^3\\
		\dot{w}^3 &=& \frac{2\lambda}{R}w^2,
	\end{array}
	\right.
	\label{eq: ode-system1}
\end{eqnarray}

\begin{eqnarray}
	\left( \frac{dR}{dt}\right) ^2 + 1=- \frac{1}{R^2}\lambda w^{1}. \label{ScaleIn}
\end{eqnarray}
We set the following conditions. The expression (\ref{eq:einstein-dirac-equation3})  must verify that at time $t=t_B, \frac{dR}{dt}|_{t_{B}}=0 \, \, \text{and } R(t_{B})=R_0$. This implies that
$  w^{1}=-\frac{R^2_0}{\lambda}$ a constant. The solution is a simplification of the big picture. The solution to the differential scale function  equation is

\begin{eqnarray}
 R(t)=\sqrt{-\left(t-R_0 \right)^2 + R^2_0}.  \label{scalefx}
\end{eqnarray}
 
 The form of this solution is confirmed by scale function for the radiation dominated Universe proposed by Plebanski and Krasinski in \cite{plebanski2006introduction} in page 289. It reads as
 \begin{eqnarray*}
 R(t) &=& \sqrt{-k\left(t - t_B - \frac{1}{k}\sqrt{\frac{\epsilon_0}{3}} \right)^2 + \frac{\epsilon_0}{3k}}, \quad \text{when } k \neq 0, \\
 R(t)&=&\sqrt{\frac{\epsilon_0}{3}(t-t_B)}, \quad \text{when } k=0,
\label{scalefx2}
\end{eqnarray*}
where $\epsilon_0$ is the radiation density, $k$ takes values $1 \; \text{or} -1$.
\begin{figure}[h]
		\includegraphics[scale=0.45, angle=0]{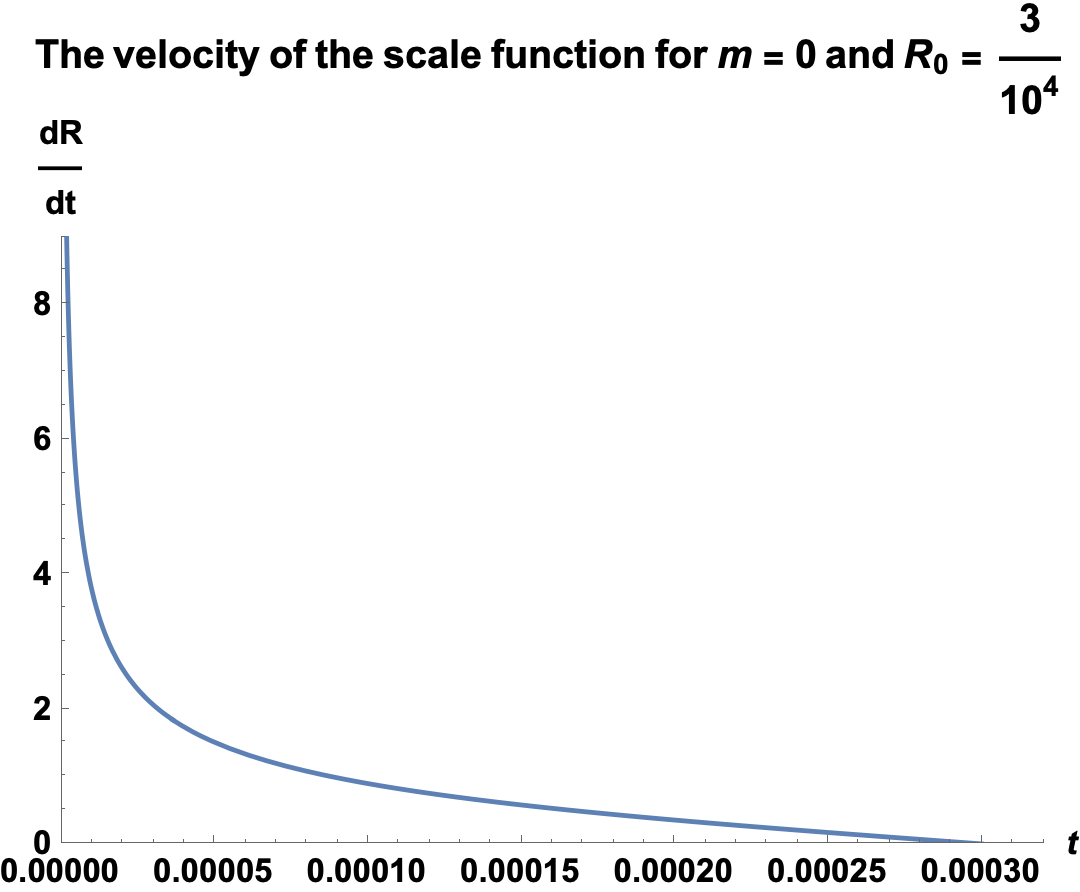}
		\centering
		\caption{  $\frac{dR(t)}{dt}$  from time $t=0$ to $t=R_0$. If one considers the today scale function is set $R=1$, the scale function being without unity, we may calibrate time so that, when the radiative forces cease to be the driving one, the corresponding time would be $t=R_0$. In the literature we know that the radiation dominated Universe epoch elapsed $\approx 72 000 years $ \cite{moore2014relativite}, page 331. }\label{scalefig1} 
	\end{figure}

Having the initial condition to the Bloch vector differential equation 
\begin{footnotesize}
\begin{equation}
\vec{w}(t_B)=\left(-\frac{R^2_0}{\lambda}, \sqrt{N^2 -\frac{R^4_0}{\lambda^2}}\sin(\phi_0), \sqrt{N^2 -\frac{R^4_0}{\lambda^2}}\cos(\phi_0) \right),
\end{equation}
\end{footnotesize}

and  substituting  $R(t)$ into (\ref{eq: ode-system1}), and based on the relationship between spherical and hyperbolic trigonometry, we can find the subsequent solutions:

\begin{eqnarray}
	\left\{
	\begin{array}{lll}
		{w}^1 &=& -\frac{R^2_0}{\lambda},\\
		{w}^2 &=&  \sqrt{N^2 -\frac{R^4_0}{\lambda^2}}\sin{\left(\Phi_{in}(t)\right)},\\
		{w}^3 &=& \sqrt{N^2 -\frac{R^4_0}{\lambda^2}} \cos{\left( \Phi_{in}(t)\right)},
	\end{array}
	\right.
	\label{eq: Sol2-ode-system1}
\end{eqnarray}\
where
\begin{equation}
\Phi_{in}(t)=4 \lambda G(t)-\lambda \pi -\phi_0 \label{fin}
\end{equation}
 $ G(t) =  \tan^{-1}\left(\frac{t}{R(t)}\right) $ and $\phi_0 $ a phase. We can notice that the Bloch vector is a function of the scale function $R(t)$.

\begin{figure}[h]
		\includegraphics[scale=0.45, angle=0]{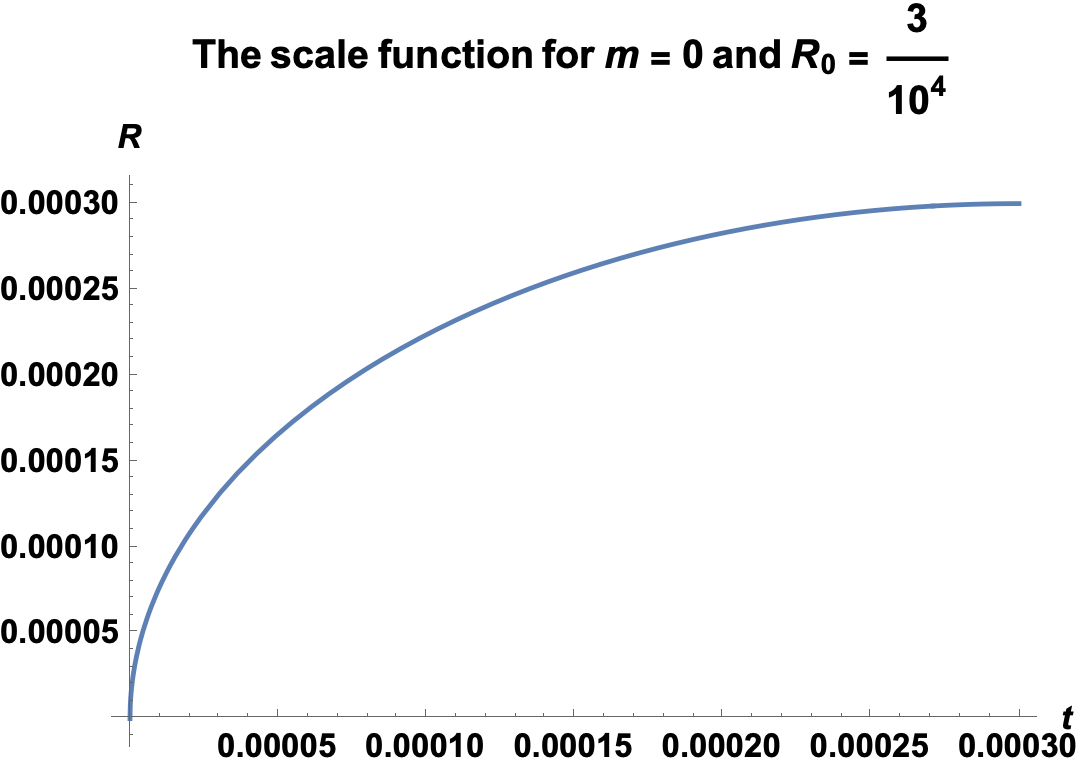}
		\centering
		\caption{ $R(t)$ increasing from time $t=0$ to $t=R_0$. }
	\end{figure}


Subtsituting $m=0$ into the components of $ \vec{w}$ in (\ref{eq:component-w1}), (\ref{eq:component-w2}) and (\ref{eq:component-w3}), we obtain

\begin{eqnarray}
	\left\{
	\begin{array}{lll}
		{w}^1 &=&  2\text{Re}\left(\alpha\bar{\beta}\right) \\
		{w}^2 &=& 2\text{Im}\left(\alpha\bar{\beta}\right)\\
		{w}^3 &=& |\alpha|^2 - |\beta|^2.
	\end{array}
	\right.
	\label{eq: Components1}
\end{eqnarray}
 Replacing $\alpha$ by $\rho_ {in_1} e^{i\eta_{in_1}}$ and $\beta$ by $\rho_ {in_2} e^{i\eta_{in_2}}$ into (\ref{eq: Components1}), where  $\rho$ is the modulus of $\alpha$ and $\eta$ its argument. We differentiate modulus and arguments from pre-selected and post-selected states with indices. We obtain
 \begin{eqnarray}
 	\left\{
 	\begin{array}{lll}
 		{w}^1 &=&  2\rho_ {in_1}   \rho_ {in_2}  \cos(\eta_{in_1}- \eta_{in_2} ), \\
 		{w}^2 &=&  2\rho_ {in_1}  \rho_ {in_2}   \sin(\eta_{in_1}- \eta_{in_2}  ),\\
 		{w}^3 &=&\rho_ {in_1}^2 - \rho_ {in_2}^2.
 	\end{array}
 	\right.
 	\label{eq: Components2}
 \end{eqnarray}
 These results constitute additional constraints for the solutions (\ref{eq: Sol2-ode-system1}). Consequently ${w}^1, 	{w}^2, 	{w}^3 $ must be real numbers.  For $ G(t)$ to be real, It has to be in the interval $] -\pi/2, \pi/2[$. 

From the expression (\ref{eq: Components2}) we obtain
{\small \begin{eqnarray}
	\eta_{in_1}- \eta_{in_2}  &=& -\arctan \left( \frac{ \sqrt{\lambda^2N^2 -R^4_0}\sin{\left(\Phi_{in}(t)\right)}}{R^2_0}\right).\label{eq: Sol3-ode-system1}
\end{eqnarray}	}

Also from (\ref{eq: Components2}) and using the normalisation condition from (\ref{eq:bloch-vector-length}), we obtain
\begin{align}
\rho_ {in_1} &=\frac{\sqrt{2}}{2} \sqrt{N  + \sqrt{N^2 -\frac{R^4_0}{\lambda^2}}\cos \left( \Phi_{in}(t)\right) }, \label{eq:component-ro1} \\
\rho_ {in_2} &= \frac{\sqrt{2}}{2}\sqrt{N  -\sqrt{N^2 -\frac{R^4_0}{\lambda^2}}\cos \left(\Phi_{in}(t)\right)  }. \label{eq:component-ro2} 
\end{align}

To sum up, $\mid \Psi_{in} \rangle $ is represented by the knowledge of the components of the three-vector $\vec{w}$  or $\alpha_{\text{in}} = \rho_ 1 e^{i\eta_1}$  and  $ \beta_{\text{in}} =\rho_ 2e^{i \eta_2 }$ at a certain $t_{in}$ chosen such that the scale function is real.

The time in the scale function is defined within the interval $[0,R_0] $ but we will consider $t_{in}$ in the interval $ –]0,R_0]$ for the Bloch vector to be well defined. Within this period, $ –]0,R_0]$, the scale function increases and stops  at $t=R_0$. But in the general picture, the driving force due to matter takes over and the Universe continues to expand. Let us point out also that the period within the radiation-dominated Universe is very short compared to the lifespan of the Universe. This period is approximated as $72000$ years vs $13,9$Gyrs today \cite{moore2014relativite}. We also remind  that within this period, the scale function is very small. The radiative period ceases where the scale function R approximately attains the value  $ R \approx 3. 10^{-4}$.   Finster in \citep{finster2011spatially} shows a bouncing behaviour of the scale function around singularity.


\subsection{$\mid \Psi_{out}\rangle$: Solution of the Einstein-Dirac equation for  $\lambda=0$. }

When $\lambda=0 $,  the ODEs in (\ref{eq:ode-system}) becomes:

\begin{eqnarray}
	\left\{
	\begin{array}{lll}
		\dot{w}^1 &=& 0,\\
		\dot{w}^2 &=& -2m{w}^3,\\
		\dot{w}^3 &=& 2m{w}^2, 
	\end{array}
	\right.
	\label{eq: ode-system3}
\end{eqnarray}

\begin{eqnarray}
	\left( \frac{dR}{dt}\right) ^2 + 1=- \frac{1}{R}m w^{1}.  \label{eq: ode-system4}
\end{eqnarray}

We differ from Finster in \citep{finster2011spatially} and we set that at $t_{max}, \frac{dR}{dt}|_{t_{max}}=0$, and $ R(t_{max})=R_{max}$. This implies that $w^{1}= - \frac{R_{max}}{m}$. One easily sees  that the two differential equations decouple. The Bloch vector solution will not depend on the scale function. These equations yield the following solutions

\begin{eqnarray}
	 R(\tau)&=&\frac{R_{max}}{2}\left( 1- \cos{(\tau)}\right),\\
	 t(\tau)&=&\frac{R_{max}}{2}\left( \tau -\sin{\left(\tau\right)} \right), \label{eq: scaleout}
\end{eqnarray}
where $\tau$ is a parameter varying between $0$ and $2\pi $. The parametric equations represent a cycloid.
\begin{figure}[h]
		\includegraphics[scale=0.45, angle=0]{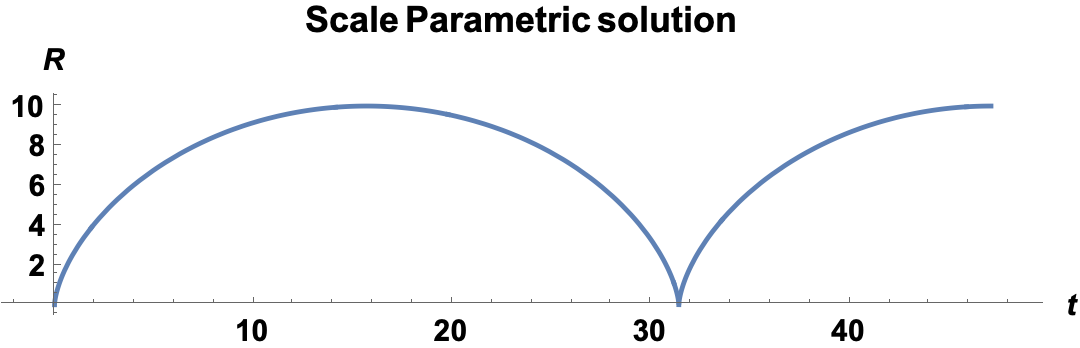}
		\centering
		\caption{ This parametric solution is set for $R_{max}=10$. This value is taken from Finster in \cite{finster2011spatially}. The time is derived from the formula $t(\tau)=\frac{R_{max}}{2}\left( \tau -\sin{(\left(\tau\right)} \right)$. }
		\label{scalfig2}
\end{figure}

The solution to the Bloch vector differential equation for the following conditions
\begin{equation}
\vec{w}(t_{\text{max}})=\left(w^1_{\text{max}}, \rho\cos(\phi_{\text{max}}), \rho\sin(\phi_{\text{max}}) \right),
\end{equation}
gives the following solution:

\begin{eqnarray}
	\left\{
	\begin{array}{lll}
		{w}^1 &=& - \frac{R_{max}}{m}\\
		{w}^2 &=& \sqrt{N^2 -\frac{R^2_{max}}{m^2}}\cos{\left(\Phi_{out}(t)\right)}\\
		{w}^3 &=& \sqrt{N^2 -\frac{R^2_{max}}{m^2}}\sin{\left(\Phi_{out}(t)\right)},
	\end{array}
	\right.
	\label{SolBlochout}
\end{eqnarray}
where
\begin{equation}
\Phi_{out}(t)=2mt + \phi_{\text{max}} - 2mt_{\text{max}}
\end{equation}\label{(fout)}
 $ m$ is  the mass of Dirac particle and $\phi_{\text{max}}$ a phase. 


Substituting $\lambda=0$ into the components of $ \vec{w}$ in (\ref{eq:component-w1}-\ref{eq:component-w3}), we obtain

\begin{eqnarray}
	\left\{
	\begin{array}{lll}
		{w}^1 &=& -\left(  |\alpha|^2 - |\beta|^2 \right) \\
		{w}^2 &=& 2\text{Im}\left(\alpha\bar{\beta}\right)\\
		{w}^3 &=&  2\text{Re}\left(\alpha\bar{\beta}\right). 
	\end{array}
	\right.
	\label{eq: Components3}
\end{eqnarray}
Replacing $\alpha$ by $ \rho_ {out_1} e^{i\eta_{out_1}}$ and $\beta$ by $ \rho_ {out_2}e^{i\eta_{out_2}}$ into (\ref{eq: Components1}), we obtain 
\begin{eqnarray}
	\left\{
	\begin{array}{lll}
		{w}^1 &=& -\left(  \rho_ {out_1}^2 - \rho_ {out_2} ^2 \right), \\
		{w}^2 &=&  2\rho_ {out_1}  	\rho_ {out_2} \sin(\eta_{out_1}- \eta_{out_2}),\\
		{w}^3 &=&  2\rho_ {out_1}  	\rho_ {out_2} \cos(\eta_{out_1}- \eta_{out_2}).
	\end{array}
	\right.
	\label{eq: Components3}
\end{eqnarray}
These results constitute additional constraints for the solutions (\ref{eq: Sol2-ode-system1}). Consequently ${w}^1,	{w}^2, 	{w}^3 $ must be real numbers.

From the expression (\ref{eq: Components3}) we obtain
 \begin{eqnarray}
\eta_{out_1}- \eta_{out_2} &=& \frac{\pi}{2} - \Phi_{out}(t) .
\end{eqnarray}	\label{Etaout}
We also obtain  from (\ref{eq: Components3})  and the norm (\ref{eq:bloch-vector-length}):
 \begin{align}
	\rho_ {out_1} &= \frac{\sqrt{2}}{2} \sqrt{N + \frac{R_{max}}{m}}, \label{eq:component-ro3} \\
	\rho_ {out_2} &=\frac{\sqrt{2}}{2} \sqrt{N- \frac{R_{max}}{m}}. \label{eq:component-ro4} 
\end{align}
 \begin{figure}[h]
		\includegraphics[scale=0.45, angle=0]{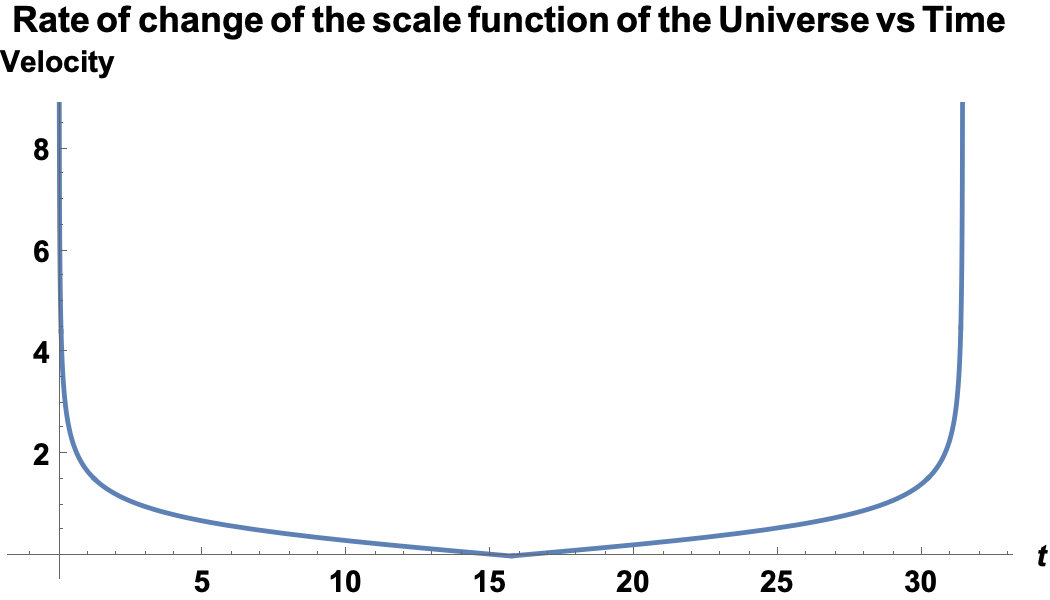}
		\centering
		\caption{ Rate of change the scale function in an Universe dominated by matter. }
		\label{scalfig2b}
\end{figure}

To sum up, $\mid \Psi_{out} \rangle $ is represented by the knowledge of the components of the three-vector $\vec{w}$  or $\alpha_{\text{out}} =  \rho_ {out_1}  e^{i\eta_{out_1}}$  and  $ \beta_{\text{out}} = \rho_ {out_2} e^{\eta_{out_2}}  $ at a certain $t_{out}$ chosen such that the scale function is real.
\section{Weak measurement on the Einstein-Dirac solutions}
A weak value of any observable is given by 
\begin{equation}
	 \langle \hat{A} \rangle_w = \frac{\langle \psi_{out}(t) |\hat{A} |\psi_{in}(t)\rangle  }{\langle \psi_{out}(t)|\psi_{in}(t)\rangle } \label{eq:weakmeasA}.
\end{equation}
wherein, $\langle \psi_{out}(t)|=\langle \psi_{out}(t_{out})|U^{\dagger}(t, t_{out})$ and \newline $|\psi_{in}(t)\rangle= U(t, t_{in})|\psi_{in}(t_{in})\rangle$.

Weak values are results of weak measurements. They are also derivable independently from weak measurements from the classical expectation values in classical quantum mechanics \cite{Shikano12}. For instance, one can derive it from the expectation values of an observable and see it as intimately linked to the weak values. If $\hat{A}$ is an observable, then
\begin{align*}
\langle \psi | \hat{A} | \psi\rangle &= \langle \psi | \hat{A} \int | \phi\rangle  \langle \phi | d\phi |\psi\rangle \\
									&=   \int \langle \psi | \hat{A}| \phi\rangle  \langle \phi | \psi\rangle d\phi \\
									&=   \int \frac{\langle \psi | \hat{A}| \phi\rangle }{\langle \psi | \phi\rangle} |  \langle \phi | \psi\rangle|^2 d\phi \\
		\langle \hat{A} \rangle		&=   \int \langle \hat{A} \rangle_w  P_{\phi \psi}d\phi 
\end{align*}
wherein $\langle \hat{A} \rangle_w=\frac{\langle \psi | \hat{A}| \phi\rangle }{\langle \psi | \phi\rangle}$ is the so called weak value of the observable $\hat{A}$ and $ P_{\phi \psi}=| \langle \phi | \psi\rangle|^2$ is the  detection probability of event $\phi$ out of the prepared initial state $\psi$. This probability is that of measuring  $\langle \hat{A} \rangle_w$.

An alternative way is to see $|\psi \rangle$ as a sum or difference of two other state vectors, let say $|\psi \rangle= |\psi_1 \rangle \pm |\psi_2 \rangle $.  Then we will have 
\begin{align*}
\langle \psi | \hat{A} | \psi\rangle &= (\langle \psi_1 | \pm \langle \psi_2 |) \hat{A} (|\psi_1 \rangle \pm |\psi_2 \rangle) \\
									&=  \langle \psi_1 | \hat{A} | \psi_1\rangle + \langle \psi_2 | \hat{A} | \psi_2 \rangle \pm 2 Re \langle \psi_1 | \hat{A} | \psi_2\rangle\\
									&=   \langle \psi_1 | \hat{A} | \psi_1\rangle + \langle \psi_2 | \hat{A} | \psi_2 \rangle \pm 2 Re \frac{\langle \psi_1 | \hat{A} | \psi_2\rangle}{ \langle \psi_1 | \psi_2\rangle } \langle \psi_2 | \psi_1\rangle\\
									&=   \langle \hat{A}_1 \rangle + \langle \hat{A}_2 \rangle \pm 2 Re \langle \hat{A}_{1,2} \rangle_w \langle \psi_2 | \psi_1\rangle
\end{align*}
wherein $ \langle \hat{A}_1 \rangle= \langle \psi_1 | \hat{A} | \psi_1\rangle $, $ \langle \hat{A}_2 \rangle= \langle \psi_2 | \hat{A} | \psi_2\rangle $ and $ \langle \hat{A}_{1,2} \rangle_w  =\frac{\langle \psi_1 | \hat{A} | \psi_2\rangle}{ \langle \psi_1 | \psi_2\rangle }  $.

\begin{figure}[h]
   \includegraphics[scale=0.65, angle=0]{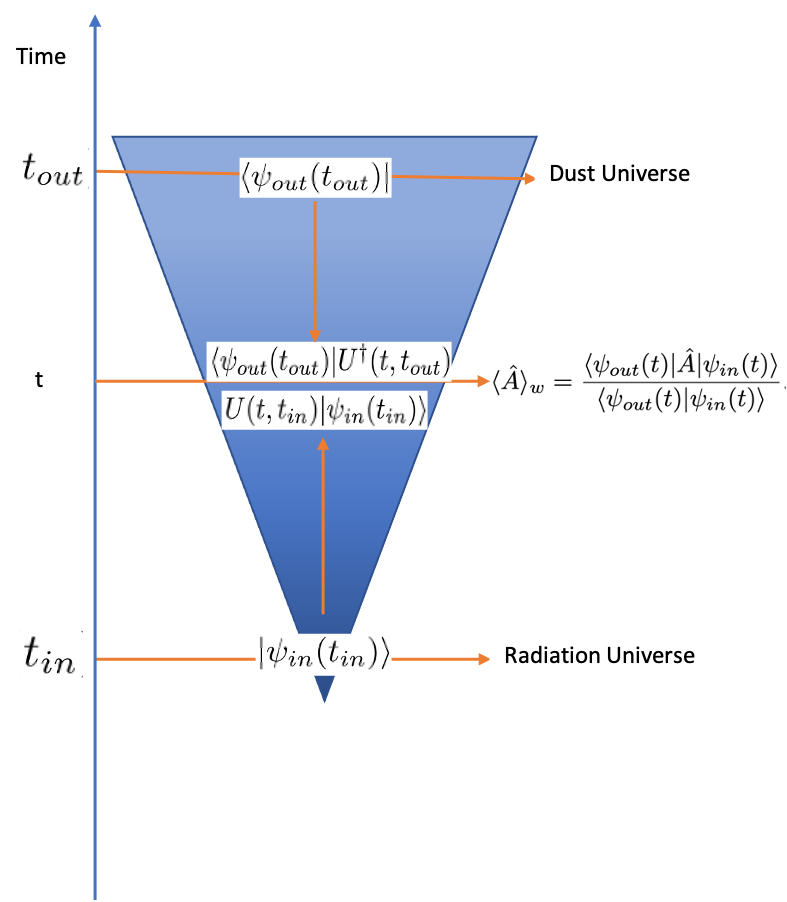}
   	\centering
   	\caption{Weak measurement at time t.}
   	\label{fig:weakfig1}
\end{figure}

In (\ref{eq:weakmeasA}), state vectors are multiplied by evolutionary operators. A pictorial description of weak measurement in our context is represented in figure \ref {fig:weakfig1}.

The solutions of Einstein-Dirac equation $|\psi_{in}(t_{in})\rangle$ and  $|\psi_{out}(t_{out})\rangle$ are defined at the respective times $t_{in}$ and $t_{out}$. We must construct the evolutionary operator to allow the weak measurement to occur at time t between  $t_{in}$ and $t_{out}$.

Because of the complexity involving the construction of the exact evolutionary operator due to the time dependency of the Dirac Hamiltonian, we are going to approximate it by using the \emph{Wentzel–Kramers–Brillouin (WKB)} method. Also, following Finster \cite{finster2009dirac}, the smallness of the Compton wavelength compared to the lifetime of the Universe justifies the use of WKB-type approximation. The approximated unitary evolutionary operator will be expressed as
\begin{equation}
U(t, t_0)=U^{-1}(t)\left( \begin{array}{cc}
	e^{-iF(t,t_0)} & 0 \\
	0 & e^{iF(t,t_0)}
\end{array}\right) U(t_0).\label{eqUt1}
\end{equation}
wherein $F(t, t_0)=-F( t_0, t)=\int ^t _{t_0} \frac{\sqrt{R^2 (t')m^2 + \lambda^2}}{R(t')}dt'$.\newline
$U(t)$ is defined such that
{\small \begin{equation}
	U(t) \begin{pmatrix}
		m & -\frac{\lambda}{R(t)} \\
		-\frac{\lambda}{R(t)} & -m
	\end{pmatrix}  U(t)^{-1}=\frac{\sqrt{R^2 (t)m^2 + \lambda^2}}{R(t)} \begin{pmatrix}
	1 & 0 \\
	0& -1
	\end{pmatrix}.\label{eqUt2}
\end{equation}}

The equation (\ref{eqUt2}) yields $g(t)=\frac{1}{2} \arctan(\frac{\lambda}{m R(t)})$ for the unitary matrix  given by
\begin{equation}
	U(t)= \begin{pmatrix}
		\cos(g(t)) & -\sin(g(t)) \\
		\sin(g(t)) & \cos(g(t))
	\end{pmatrix}.  
\end{equation}

The computation of $\left( \begin{array}{cc}
	e^{-iF(t,t_0)} & 0 \\
	0 & e^{iF(t,t_0)}
\end{array}\right) $ follows Finster in \cite{finster2009dirac}.

The unitary operator applied to the pre-selected state
$U(t, t_{\text{in}} ) $ is given as

 $$U^{-1}(t)\left( \begin{array}{cc}
	e^{-iF(t ,t_{\text{in}} )} & 0 \\
	0 & e^{iF(t,t_{\text{in}} )}
\end{array}\right)U(t_{in}).$$ 
Since the pre-selected state is taken from the radiation-dominated Universe where $m=0$, this will yield\\ $g(t_{\text{in}})= \frac{\pi}{4} + k \pi$.  For $k=0$, 
\begin{equation}
g(t_{\text{in}})=\frac{\pi}{4} \label{piin}
\end{equation}
 and the matrix will be given by
\begin{equation}
U(t_{\text{in}})= \frac{\sqrt{2}}{2}\begin{pmatrix}
	1 & -1 \\
	1 & 1
\end{pmatrix}. \label{matrin}
\end{equation}

Whereas the post-selected state is taken from the dust-dominated Universe wherein $\lambda=0$. This will yield \\   $g(t_{out})=0 + k\pi$  for $\lambda=0 $, and for $k=0$ 
\begin{equation}
g(t_{\text{in}})=0 
\end{equation}\label{piout}
and the unitary matrix $U(t_{out})$ is the identity matrix 
\begin{equation}U(t_{out})= \left( \begin{array}{cc}
	1 & 0 \\
	0 & 1
\end{array}\right).\end{equation}\label{matrout}.

Knowing that $\left. U^{\dagger}(t, t_{\text{out}})=U(t_{\text{out}}, t)\right.$, this later is given by 
$$U^{-1}(t_{\text{out}})\left( \begin{array}{cc}
	e^{-iF(t _{\text{out}},t)} & 0 \\
	0 & e^{iF(t_{\text{out}},t)}
\end{array}\right)U(t).$$ 

The unitary operator approximated, we can now compute the weak value of some observables starting with the energy-momentum tensor.

\subsection{Weak measurement of the Energy-momentum tensor. }

In this section, we compute the weak value of the energy-momentum tensor given by

\begin{equation}
	\langle  T_{\mu\nu} \rangle_w = \frac{\langle \psi_{out}(t) | T_{\mu\nu} |\psi_{in}(t)\rangle  }{\langle \psi_{out}(t)|\psi_{in}(t)\rangle } \label{eq:weakmeasT}.
\end{equation}

More explicitly, we have 
\begin{equation}
	\langle  T_{\mu\nu} \rangle_w = \frac{1}{2} \frac{\langle \psi_{\text{out}}(t) | \left( i G_{\mu} D_{\nu}  + i G_{\nu} D_{\mu} \right)|\psi_{\text{in}}(t)\rangle}{\langle \psi_{\text{out}}(t)|\psi_{\text{in}}(t)\rangle } \label{eq:weakmeasTexp}.
\end{equation}

where $G^{\mu}  $ are the linear combinations of the Dirac matrices of Minkowski space given by:
{\small
\begin{align}
	& G^{0} = \gamma^{0}, \\
	& G^{r} = \frac{f(r)}{R(t)} \left( \cos{\theta }\,\gamma^{3} + \sin{\theta }\cos{\phi}\, \gamma^{1} + \sin{\theta} \sin{\phi}\, \gamma^{2} \right), \\
	& G^{\theta} = \frac{1}{r R(t)} \left( -\sin{ \theta}\, \gamma^{3} + \cos{ \theta} \cos{ \phi }\,\gamma^{1} + \cos{ \theta} \sin{ \phi}\, \gamma^{2} \right), \\
	& G^{\phi} = \frac{1}{r R(t) \sin{ \theta}} \left( -\sin{ \phi} \,\gamma^{1} + \cos{ \phi} \,\gamma^{2} \right),
\end{align}
}
with $f(r)=\sqrt{1-r^2}$.  Note that $G_{\mu}=g_{\mu\nu}G^{\nu}  $.

The Dirac matrices are given as
\[
\gamma^0 = \begin{pmatrix} I & 0 \\ 0 & -I \end{pmatrix}, \quad
\gamma^{\alpha} = \begin{pmatrix} 0 & \sigma^{\alpha} \\ -\sigma^{\alpha} & 0 \end{pmatrix},
\]

where $I$ is the $2 \times 2$ identity matrix and $\sigma^{\alpha}$ are the Pauli matrices define in(\ref{paulim}).

The differential operator is given by $D_{\mu} =\partial_{\mu} -i E_{\mu}  \text {, }$ and  $G^{\mu} $  obeys to  the anti-commutation rule
$$
\left\{G^{\mu} , G^{\nu} \right\} \equiv G^{\mu}  G^{\nu} +G^{\nu} G^{\mu} =2 g^{\mu \nu} \mathbf{1}_{4} .
$$
The so-called spin coefficients are given by $E_{\mu}$.

For an orthogonal metric, the combination $G^{\mu} E_{\mu}$ takes the simple form (for more details see  \cite{finster2011spatially})

$$
G^{\mu} E_{\mu}=\frac{i}{2 \sqrt{|g|}} \partial_{\mu}\left(\sqrt{|g|} G^{\mu}\right) \quad \text { with } \quad g=\operatorname{det} g_{i j},
$$

making it unnecessary to compute the spin connection coefficients or even the Christoffel symbols.

To compute the numerator part of the weak value of the energy-momentum tensor, following Finster in\cite{finster2009dirac}, we adapt the object $P(t, \vec{x}; t', \vec{x'}) $ to our need

{\footnotesize
\begin{equation}
	= \left(R(t) R(t')\right)^{-\frac{3}{2}} E_{\lambda}(\vec{x}, \vec{x'}) \otimes
	\begin{pmatrix}
		 \alpha_{\text{in}}(t) \overline{\alpha_{\text{out}}(t')}  -\alpha_{\text{in}}(t) \overline{\beta_{\text{out}}(t')} \\
		\beta_{\text{in}}(t) \overline{\alpha_{\text{out}}(t') }    -\beta_{\text{in}}(t) \overline{\beta_{\text{out}}(t')}
	\end{pmatrix}.
\end{equation}
}

wherein $E_{\lambda}$ denote the spectral projectors of  $\mathcal{D}_{S^{3}}$, given by  $E_{\pm \mid \lambda \mid} =\displaystyle  \sum^{\mid \lambda \mid - \frac{3}{2}}_{n=0}\sum^{j}_{k=-j}\psi^{\pm}_{njk}(x) \overline{\psi^{\pm}_{njk}(\acute{x})}$. Then, the numerator part of weak value of the energy-momentum tensor, $ \langle \psi_{\text{out}} | T_{\mu\nu} | \psi_{\text{in}} \rangle(t, \vec{x})$ can be expressed in terms of $P$ by

\begin{equation}
	\begin{split}
		 = \frac{1}{2} \operatorname{Tr}_{\mathbb{C}^{4}} \Big\{ &\left(i G_{\mu} D_{\nu} + i G_{\nu} D_{\mu}\right) 
		 P\left(t, \vec{x} ; t^{\prime}, \vec{x}^{\prime}\right) \Big\}\Bigg|_{t^{\prime}=t, \vec{x}^{\prime}=\vec{x}}.
	\end{split}
\end{equation}

Because of the homogeneity and the isotropy of the FLRW space, only the diagonal energy-momentum tensor components are different from 0. We get the following results after computations and normalization

%
%
%

\begin{widetext}

\begin{equation}
	\begin{aligned}
& \langle  T_{0}^{0} \rangle_w=\frac{\langle \psi_{out}(t)| T_{0}^{0} |\psi_{in}(t)\rangle }{\langle \psi_{out}(t) | \psi_{in}(t)\rangle} =\frac{\Bigl[ m R(t) \left(\alpha_{\text{in}}(t) \overline{\alpha_{\text{out}}}(t) - \beta_{\text{in}} (t)\overline{\beta_{\text{out}}}(t)\right) -\lambda \left(   \alpha_{\text{in}(t)} \overline{\beta_{\text{out}}}(t) +  \beta_{\text{in}}(t) \overline{\alpha_{\text{out}}}(t)  \right) \Bigr] }{R(t)\left(\overline{\alpha_{\text{out}}}(t) \alpha_{\text{in}}(t) + \overline{\beta_{\text{out}}}(t) \beta_{\text{in}}(t) \right) }, 
\end{aligned}
\end{equation}

and 

	\begin{equation}
		\begin{aligned}
			&
			\langle  T_{j}^{j} \rangle_w= \frac{\langle \psi_{out}(t)| T_{j}^{j} |\psi_{in}(t)\rangle }{\langle \psi_{out}(t) | \psi_{in}(t)\rangle} =  \frac{\lambda \left(   \alpha_{\text{in}}(t) \overline{\beta_{\text{out}}}(t) +  \beta_{\text{in}}(t) \overline{\alpha_{\text{out}}}(t)  \right) }{3R(t)\left(\overline{\alpha_{\text{out}}}(t) \alpha_{\text{in}}(t) + \overline{\beta_{\text{out}}}(t)\beta_{\text{in}} (t)\right)},\\
		\end{aligned}
	\end{equation}
\end{widetext}
where j represents $r, \theta, \phi$. The space diagonal energy momentum tensor are all equal, that is  $T_{r}^{r}=T_{\theta}^{\theta}=T_{\phi}^{\phi}$.

One sees here that if $|\psi_{in}(t)\rangle =|\psi_{out}(t)\rangle$, the weak value of the energy-momentum tensor will coincide with the one computed in \citep{finster2011spatially}. In this regard, weak measurements are generalizations of classical averages.

Another quick observation is that if 
 \[\overline{\alpha_{\text{out}}}(t)\alpha_{\text{in}}(t) + \overline{\beta_{\text{out}}}(t)\beta_{\text{in}} (t) \approx 0,\] an amplification phenomenon occurs. This is realized when $\overline{\alpha_{\text{out}}}(t) \approx -k\beta_{\text{in}}(t)$ and $\overline{\beta_{\text{out}}}(t) \approx k\alpha_{\text{in}}(t)$, where $k$ is a factor or a phase.

For better analysis, we construct a \emph{complex Bloch vector}. Aharonov and Gruss constructed a density operator from the Two-State vectors in their paper \emph{Two-time interpretations} of quantum mechanic \cite{aharonov2005two}. This complex Bloch vector is linked to that density operator. One should note that if the pre-selected state is equal to the post-selected one, then we shall get the same construction as in the paper of Finster and Hainzl on page 14 in \citep{finster2011spatially}. Our complex Bloch vector will be given as
\[
\vec{v}_{f,i} =
\begin{pmatrix}
	v_{f,i}^1 \\
	v_{f,i}^2 \\
	v_{f,i}^3
\end{pmatrix} =
\begin{pmatrix}
	\langle \xi_{\text{out}}(t) | \sigma^1 | \xi_{\text{in}}(t) \rangle \\
	\langle \xi_{\text{out}}(t) | \sigma^2 | \xi_{\text{in}}(t) \rangle \\
	\langle \xi_{\text{out}}(t) | \sigma^3 | \xi_{\text{in}}(t) \rangle
\end{pmatrix}
\]

wherein  $\xi_{\text{out}} (t) = \left( \begin{array}{c}
	\alpha_{\text{out}} (t)	\\
	\beta_{\text{out}}(t)
\end{array}\right)$ and  $\xi_{\text{in}} (t) = \left( \begin{array}{c}
	\alpha_{\text{in}} (t)	\\
	\beta_{\text{in}}(t)
\end{array}\right)$.

We also define  $v_{f,i}^0$ as $\langle \xi_{\text{out}}(t) | I_{2\times 2} | \xi_{\text{in}}(t) \rangle $ where $I_{2\times 2} $ is the 2 dimensional identity matrix. This will yield $\overline{\alpha_{\text{out}}}(t) \alpha_{\text{in}}(t) + \overline{\beta_{\text{out}}}(t)\beta_{\text{in}} (t)$. \\
We can now redefine the weak energy-momentum tensor as
\begin{equation}
	\begin{split}
&\langle  T_{0}^{0} \rangle_w= \frac{\Bigl[ m R(t) \left( v_{f,i}^3 \right) -\lambda \left( v_{f,i}^1 \right) \Bigr] }{R(t)\left(v_{f,i}^0\right) } \\
&\langle  T_{j}^{j} \rangle_w =  \frac{\lambda \left(   v_{f,i}^1 \right) }{3R(t)\left( v_{f,i}^0 \right)}
	\end{split}
\end{equation}
 
 	Let us express the energy-momentum tensor weak values in \textbf{ terms of $\alpha_{\text{in}}(t_{\text{in}})$ and $\alpha_{\text{out}}(t_{\text{out}})$} wherein  \\ $A= e^{i\left(F(t_{\text{out}}, t_{\text{in}}) \right)}$,  $\bar{A}= e^{-i\left(F(t_{\text{out}}, t_{\text{in}}) \right)}$, \\$B= e^{i\left( F(t, t_{\text{in}}) + F(t, t_{\text{out}}) \right)} $ and $\bar{B}= e^{-i\left( F(t, t_{\text{in}}) + F(t, t_{\text{out}}) \right)}$. Because of the lengthiness  of the expression, we break it as follows
 	
 	\begin{widetext}
 				\begin{equation} \begin{split}
						&	v_{f,i}^0=\frac{\sqrt{2}}{2}\left(\overline{\alpha_{\text{out}}}(t_{\text{out}}) \alpha_{\text{in}}(t_{\text{in}}) - \overline{\alpha_{\text{out}}}(t_{\text{out}}) \beta_{\text{in}}(t_{\text{in}})\right) \bar{A} 
						+	\frac{\sqrt{2}}{2}\left(\overline{\beta_{\text{out}}}(t_{\text{out}}) \beta_{\text{in}}(t_{\text{in}}) + \overline{\beta_{\text{out}}}(t_{\text{out}}) \alpha_{\text{in}}(t_{\text{in}})\right) A,
					\end{split}
				\end{equation}

			\begin{equation}
					\begin{split}
						&v_{f,i}^3=\frac{\sqrt{2}}{2} \alpha_{\text{in}}(t_{\text{in}}) \overline{\alpha_{\text{out}}}(t_{\text{out}}) \left[\sin{(2 g(t))} B + \cos{(2 g(t))} \bar{A} \right] 
						+ \frac{\sqrt{2}}{2}\beta_{\text{in}}(t_{\text{in}}) \overline{\beta_{\text{out}}}(t_{\text{out}}) \left[-\sin{(2 g(t))} \bar{B}- \cos{(2 g(t))} A\right] \\
						&+ \frac{\sqrt{2}}{2}\alpha_{\text{in}}(t_{\text{in}}) \overline{\beta_{\text{out}}} (t_{\text{out}})\left[\sin{(2 g(t))} \bar{B} - \cos{(2 g(t))} A\right] 
						+ \frac{\sqrt{2}}{2} \beta_{\text{in}}(t_{\text{in}}) \overline{\alpha_{\text{out}}} (t_{\text{out}})\left[\sin{(2 g(t))} B - \cos{(2 g(t))} \bar{A} \right],
					\end{split}
				\end{equation}

				\begin{equation}
					\begin{split}
						& v_{f,i}^1=\frac{\sqrt{2}}{2}\alpha_{\text{in}}(t_{\text{in}}) \overline{\alpha_{\text{out}}}(t_{\text{out}}) \left[- \sin{(2 g(t))} \bar{A}+\cos{(2 g(t))}B\right] 
						+ \frac{\sqrt{2}}{2}\beta_{\text{in}}(t_{\text{in}}) \overline{\beta_{\text{out}}}(t_{\text{out}}) \left[ \sin{(2 g(t))} A-\cos{(2 g(t))} \bar{B} \right] \\
						&+ \frac{\sqrt{2}}{2}\alpha_{\text{in}}(t_{\text{in}}) \overline{\beta_{\text{out}}} (t_{\text{out}})\left[\sin{(2 g(t))} A+ \cos{(2 g(t))}\bar{B}  \right] 
						+\frac{\sqrt{2}}{2} \beta_{\text{in}}(t_{\text{in}}) \overline{\alpha_{\text{out}}} (t_{\text{out}})\left[\sin{(2 g(t))} \bar{A} + \cos{(2 g(t))} B \right].
					\end{split}
				\end{equation}	
	\end{widetext}

 We can deduce the orthogonality condition for $v_{f,i}^0 \approx 0$. This is obtained if $\overline{\alpha_{\text{out}}}(t_{\text{out}}) \approx $:
 \begin{footnotesize}
 \begin{equation}
\begin{split}
& -\frac{\sqrt{2}}{2}\alpha_{\text{in}}(t_{\text{in}}) \left[ -\sin{(g(t_{\text{out}}))} e^{-i F(t_{\text{out}}, t_{\text{in}})}  + \cos{(g(t_{\text{out}}))} e^{i F(t_{\text{out}}, t_{\text{in}})} \right] \\
& -\frac{\sqrt{2}}{2}\beta_{\text{in}}(t_{\text{in}}) \left[ \sin{(g(t_{\text{out}}))} e^{-i F(t_{\text{out}}, t_{\text{in}})} + \cos{(g(t_{\text{out}}))} e^{i F(t_{\text{out}}, t_{\text{in}})} \right].
\end{split}
\end{equation}
 \end{footnotesize}

and $\overline{\beta_{\text{out}}}(t_{\text{out}}) \approx $
\begin{footnotesize}

 \begin{equation}
\begin{split}
& \frac{\sqrt{2}}{2}\alpha_{\text{in}}(t_{\text{in}}) \left[ \cos{(g(t_{\text{out}}))} e^{-i F(t_{\text{out}}, t_{\text{in}})}  + \sin{(g(t_{\text{out}}))} e^{i F(t_{\text{out}}, t_{\text{in}})} \right] \\
& +\frac{\sqrt{2}}{2}\beta_{\text{in}}(t_{\text{in}}) \left[ -\cos{(g(t_{\text{out}}))} e^{-i F(t_{\text{out}}, t_{\text{in}})} + \sin{(g(t_{\text{out}}))} e^{i F(t_{\text{out}}, t_{\text{in}})} \right].
\end{split}
\end{equation} 

\end{footnotesize}
 
 As it is difficult to have a simple expression of the weak value of the energy-momentum tensor to make an interpretation, let us calculate a concrete example using values that simplify the computation. 
 For the pre-selected states  let us have
\begin{equation}
	\begin{cases}
	\lambda=3/2, \text{the eigenvalue of }\mathcal{D}_{\mathcal{H}},\\
	 g(t_{\text{in}})=\frac{\pi}{4}, \,\text{see  (\ref{piin})},\\
	 U(t_{\text{in}})= \frac{\sqrt{2}}{2}\begin{pmatrix}
	1 & -1 \\
	1 & 1
\end{pmatrix},\text{see (\ref{matrin}}) \\
 R_0\,  \text{is of order}\, 10^{-4}\, ,\text{we set } R^4_0 \approx 0.
	\end{cases}
\end{equation} 
 These values will yield 
 \begin{equation}
	\begin{cases}
	\Phi_{in}(t)=3G(t)-\frac{\pi}{4}  -\phi_0, \, \text{see (\ref{fin})},\\
	\eta_{in_1}- \eta_{in_2}  = -\arctan{ \left( \frac{ 3\sin{\left(\Phi_{in}(t)\right)}}{R^2_0}\right)}, \text{see (\ref{eq: Sol3-ode-system1})} \\
	\text{we set}\, \eta_{in_2}=0\\
	\rho_ {in_1} =\frac{\sqrt{2}}{2} \sqrt{2  + 2\cos \left( \Phi_{in}(t)\right) }, \, \text{see (\ref{eq:component-ro1})},\\
	\rho_ {in_2} =\frac{\sqrt{2}}{2} \sqrt{2  - 2\cos \left( \Phi_{in}(t)\right) }, \, \text{see  (\ref{eq:component-ro2})}.
	\end{cases}
\end{equation} 

Therefore the pre-selected state at $t_{\text{in}}$ will be given by
		\begin{equation}
	\begin{cases}
	\alpha_{\text{in}}(t_{\text{in}})=\frac{\sqrt{2}}{2} \sqrt{2  + 2\cos \left( \Phi_{in}(t_{\text{in}})\right) }e^{-i\arctan{ \left( \frac{ 3\sin{\left(\Phi_{in}(t_{\text{in}})\right)}}{R^2_0}\right)}} \\
	\beta_{\text{in}}(t_{\text{in}}) =\frac{\sqrt{2}}{2} \sqrt{2  - 2\cos \left( \Phi_{in}(t_{\text{in}})\right) }
	\end{cases}
\end{equation}	
As for the post-selected state vector, the scale function and the time expressed as $R(\tau)=\frac{R_{\text{max}}}{2}\left( 1- \cos(\tau)\right)$, $t(\tau)=\frac{R_{\text{max}}}{2}\left( \tau - \sin(\tau) \right)$ respectively, we have the following observations 
\begin{itemize}
			\item A maximal $R(\tau)$ will be $R_{\text{max}}$, and the maximal $t$, that is $t_{\text{out}}$ will be $R_{\text{max}} k \frac{\pi}{2}$.
			\item We shall have: $\rho_{\text{out}1} = \sqrt{2 +\frac{R_{\text{max}}}{m}}$, \qquad \qquad $\rho_{\text{out}2}=\sqrt{2 -\frac{R_{\text{max}}}{m}}$ and $\eta_{\text{out}1} - \eta_{\text{out}2} = \frac{\pi}{2} -\phi_{\text{max}}$.
		 \end{itemize}
	We remind that  the post-selected state is taken from the dust-dominated Universe wherein $\lambda=0$. This will yield    $g(t_{out})=\frac{1}{2} \arctan(\frac{\lambda}{m* R(t_{\text{out}})})=0 + k\pi$ . This will lead,  for $k=0$, to $U(t_{out})= \left( \begin{array}{cc}
	1 & 0 \\
	0 & 1
\end{array}\right)$,  an identity matrix.

The post-selected state $t_{\text{out}}$ will be
	 \begin{equation}
		\begin{split}
			&\alpha_{\text{out}} (t_{\text{out}})=\sqrt{2 +\frac{R_{\text{max}}}{m}}e^{i\left(\frac{\pi}{2} -\phi_{\text{max}} \right)},\\
			&\beta_{\text{out}} (t_{\text{out}})=	\sqrt{2 -\frac{R_{\text{max}}}{m}}.
		\end{split}
	\end{equation}
	
For simplicity, let us consider a weak measurement taking place at time $t_{out}$  and $R_{\text{max}}=10$ and $m=21.5$. This implies that 
$$\left.  \left( \begin{array}{c}
	\alpha_{\text{in}} (t_{\text{out}})	\\
	\beta_{\text{in}}(t_{\text{out}})
\end{array}\right)=U(t_{\text{out}}, t_{\text{in}} ) \left( \begin{array}{c}
	\alpha_{\text{in}} (t_{\text{in}})	\\
	\beta_{\text{in}}(t_{\text{in}})
\end{array}\right) \right., $$

where
$$U(t_{\text{out}}, t_{\text{in}} )=\frac{1}{\sqrt{2}}\left(
\begin{array}{cc}
 e^{-i \pi  m R_\text{max}} & -e^{-i \pi  m R_\text{max}} \\
 e^{i \pi  m R_\text{max}} & e^{i \pi  m R_\text{max}} \\
\end{array}
\right),$$
and $U(t_{\text{out}}, t_{\text{out}} )= I $. $ F(t_{\text{out}}, t_{\text{in}} )= mt_{\text{out}}$. These elements yield the following state vectors 
\begin{equation}
	\begin{split}
		&\alpha_{\text{in}} (t_{\text{out}})=	\frac{\sqrt{2}}{2}(\alpha_{\text{in}}(t_{\text{in}})-\beta_{\text{in}}(t_{\text{in}})) e^{-i \pi  m R_\text{max}},\\
		&\beta_{\text{in}} (t_{\text{out}})=	\frac{\sqrt{2}}{2}(\alpha_{\text{in}}(t_{\text{in}})+\beta_{\text{in}}(t_{\text{in}})) e^{i \pi  m R_\text{max}}.
	\end{split}
\end{equation}

Since the weak measurement is taking place at time $t_{\text{out}} $, the post-selected state is fixed. 
	
 The values in the weak measurement of the energy-momentum tensor in the following expression
\begin{equation}
	\begin{split}
&\langle  T_{0}^{0} \rangle_w= \frac{\Bigl[ m R(t_{\text{out}}) \left( v_{f,i}^3 \right) -\lambda \left( v_{f,i}^1 \right) \Bigr] }{R(t_{\text{out}})\left(v_{f,i}^0\right) } \\
&\langle  T_{j}^{j} \rangle_w =  \frac{\lambda \left(   v_{f,i}^1 \right) }{3R(t_{\text{out}})\left( v_{f,i}^0 \right)}
	\end{split}
\end{equation}
will be:

\begin{equation}
\begin{split}
 &v_{f,i}^0=\text{C} \left( (A-B) \cos{(\frac{\Phi_{in}}{2})}\text{D}+(A+B) \sin{(\frac{\Phi_{in}}{2})}\right)\\   
 &v_{f,i}^1 =\text{C} (A B+1) \left( \cos{(\frac{\Phi_{in}}{2})}D-\sin{(\frac{\Phi_{in}}{2})}\right)\\
 &  v_{f,i}^3= C\left(\left(A+B\right)\cos{(\frac{\Phi_{in}}{2})} D+\left(A-B\right)\sin{(\frac{\Phi_{in}}{2})}\right)
\end{split}
\end{equation}
wherein we have set $ A=e^{2 i \pi  m \text{Rmax}} $ ,  $B=e^{i \text{$\phi $max}}$ and $C=e^{-i \pi  m \text{Rmax}} \sqrt{\frac{\text{Rmax}}{m}+2} $, $D=e^{i \tan ^{-1}\left(\frac{3 \sin (\text{$\phi $tin})}{\text{R0}^2}\right)} $.

If we set $\Phi_{in}$ such that $\Phi_{in}=\pi$, we will have
\begin{equation}
\begin{split}
 &v_{f,i}^0=\text{C} (A+B) \\   
 &v_{f,i}^1 =-\text{C} (A B+1) \\
 &  v_{f,i}^3= C\left(A-B\right)
\end{split}
\end{equation}
Such configuration happens if $t_{\text{in}}=t_B $ and $\phi_0=\pi$.

 What we can conclude is that weak measurements reveal unusual values, such as complex numbers. In our case, for  fixed $m$ and $  R_{\text{max}}$, the numerical result will depend on $\phi_{max}$.
 
In (\ref{eq:einstein-dirac-equationa2}), the acceleration of the Universe is given in terms of the energy momentum tensor. In (\ref{eq:einstein-dirac-equation3}), it is given in terms of the Bloch vector. But in the \emph{Two-State Formalism}, the Hopf transformation yields a complex Bloch vector. The \emph{Two-State Formalism} generalizes the \emph{One-State Formalism}  since it suffices to consider the same state vector in the \emph{Two-State Formalism} to have the classical Bloch vector. We propose to express the acceleration of the universe in terms of weak values of the energy-momentum tensors and derive it in terms of Complex Bloch vector. The acceleration of the Universe will then be given as

\begin{equation}
\begin{split}
    \ddot{R} & = \frac{R}{2} \left(3 \langle  T_{j}^{j} \rangle_w - \langle  T_{0}^{0} \rangle_w\right),  \\
    \ddot{R} & = \frac{\Bigl[- m R \left( v_{f,i}^3 \right) + 2\lambda \left( v_{f,i}^1 \right) \Bigr] }{2\left(v_{f,i}^0\right) }. \label{eq:einstein-dirac-equation4}
\end{split}
\end{equation}

In this formula, one clearly sees  that the acceleration of the Universe may be comprehended from the \emph{Two-State Formalism} and weak measurements theories. Assuming a different post-selected state vector may change the shape of the acceleration of the Universe. We remind here that the real and the imaginary parts of a weak value can be measured. The real part of a weak value may be interpreted as the best estimation of the conditioned average associated with an observable in Two Vector Formalism, \citep{dressel2014colloquium}, \cite{cormann2016revealing}, and they represent the shift of the average detected position due to post-selection. The imaginary part of the weak value represents the shift of the average impulsion due to post-selection. 

The computation of the acceleration of the Universe with the above condition yields
\begin{equation}
\begin{split}
		&\ddot{R}  = Re \left( \frac{\Bigl[- m R \left( v_{f,i}^3 \right) + 2\lambda \left( v_{f,i}^1 \right) \Bigr] }{2\left(v_{f,i}^0\right) } \right) \label{eq:einstein-dirac-equation4}
\end{split}
\end{equation}

This example shows that the acceleration is not Zero and depends here on the post-selection.

\begin{figure}[h!]
   \includegraphics[scale=0.35 , angle=0]{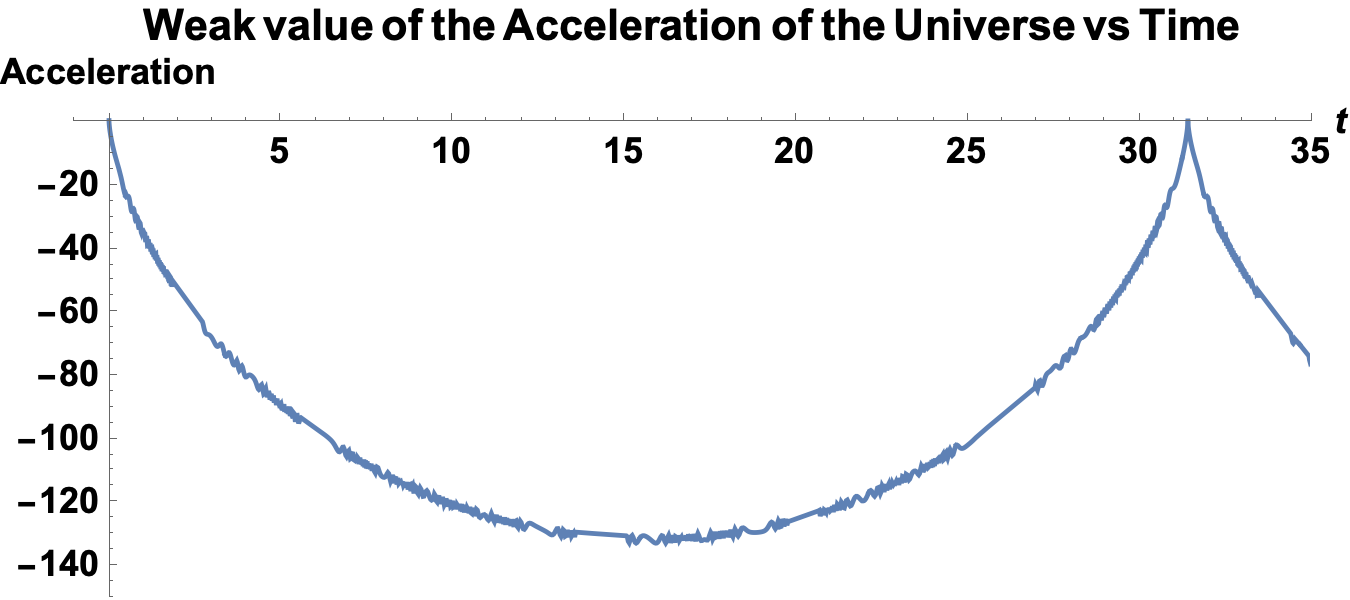}
   	\centering
   	\caption{The real part of the weak values of the acceleration of the Universe in function of time for $m=21.5, R_{\text{max}}=10, \phi_{\text{max}}=0$ and  $\lambda=\frac{3}{2}$ }
   	\label{fig:weakfig7}
\end{figure}
Figure (\ref{fig:weakfig7}) is about the weak measurement of the acceleration of the Universe as a function of time in a matter Dominated Universe.

\begin{figure}[h!]
   \includegraphics[scale=0.35 , angle=0]{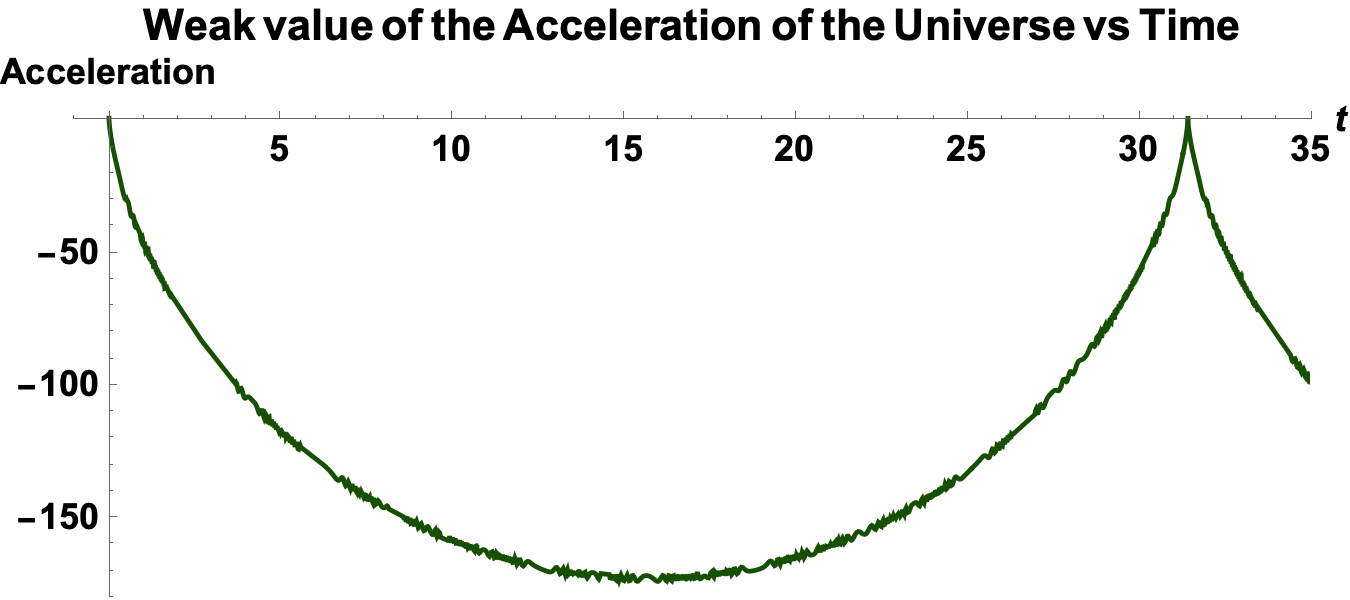}
   	\centering
   	\caption{The real part of the weak values of the acceleration of the Universe in function of time for $m=21, R_{\text{max}}=10, \phi_{\text{max}}=\frac{\pi}{4}$ and $\lambda=\frac{3}{2}$.  }
   	\label{fig:weakfig8}
\end{figure}
The difference between Figure \ref{fig:weakfig7} and  \ref{fig:weakfig8} shows how the acceleration of the Universe may change with a different post-selection. We have post-selected a state vector with $\phi_{\text{max}}=0$ in Figure \ref{fig:weakfig7} and $\phi_{\text{max}}=\frac{\pi}{4}$ in Figure \ref{fig:weakfig8}. It is clear that, this difference causes more acceleration in Figure \ref{fig:weakfig8} than in Figure \ref{fig:weakfig7}.
In conclusion, we have shown how to compute the weak value of the Energy-momentum tensor. We derived how to amplify weak values of the energy-momentum tensor using an appropriate state vector almost orthogonal to the pre-selected one. We have finally derived the weak values of the acceleration of the Universe and have showed how sensitive it is compared to classical acceleration.

\subsection{Weak value of $\sigma^z$ operator. }

In this section, we delve into the weak measurement of the $\sigma^z$ operator which is given by the following expression

\begin{equation}
	\sigma^z_{w} = \frac{\langle \psi_{out}(t)| \sigma^z |\psi_{in}(t)\rangle}{\langle \psi_{out}(t) | \psi_{in}(t)\rangle}. \label{eq:weakmeas1}
\end{equation}

Following Ferraz, Kofman and Cormann   in \cite{ferraz2022geometrical, kofman2012nonperturbative, cormann2017geometric} respectively, the weak value of the generator of $SU(2)$ is given by 

\begin{equation}
	\sigma _{r,w} =  \frac{ \vec{f}. \vec{r} + \vec{r}. \vec{i} + i\vec{f}.(\vec{r} \wedge \vec{i})}{\frac{1}{2} (1 + \vec{f} .\vec{i} )}.    \label{eq:weakmeas6}
\end{equation}

In our case, $\vec{r}=(\begin{array}{ccc}	0 & 0 & 1\end{array})$.  
Following Kofman  in  \cite{kofman2012nonperturbative}, the vectors $ \vec{f}$ and $ \vec{i}$ are computed as follows

$$\vec{w_f}= 
\begin{pmatrix}
	\langle \xi_{\text{out}} (t)| \sigma^1 | \xi_{\text{out}}(t) \rangle \\
	\langle \xi_{\text{out}}(t) | \sigma^2 | \xi_{\text{out}}(t) \rangle \\
	\langle \xi_{\text{out}}(t) | \sigma^3 | \xi_{\text{out}}(t) \rangle
	\end{pmatrix},
$$
where $\xi_{\text{out}} (t) = \left( \begin{array}{c}
	\alpha_{\text{out}} (t)	\\
	\beta_{\text{out}}(t)
\end{array}\right)$ is expressed as\\
 \begin{equation}
		\begin{split}
		\overline{	\alpha_{\text{out}}} (t)=&\overline{\alpha_{\text{out}}} (t_{\text{out}})  \cos{(g(t))} e^{-i F(t_{\text{out}}, t)} +\\
		 & \overline{ \beta_{\text{out}}} (t_{\text{out}})\sin{( g(t))} e^{i F(t_{\text{out}}, t)},
		\end{split}
\end{equation} 
and \\
 \begin{equation}
		\begin{split}
			 \overline{\beta_{\text{out}}}(t)=&-\overline{\alpha_{\text{out}}} (t_{\text{out}})  \sin{(g(t))} e^{-i F(t_{\text{out}}, t)}  +\\
			 & \overline{ \beta_{\text{out}}} (t_{\text{out}})\cos{( g(t))} e^{i F(t_{\text{out}}, t)}.
		\end{split}
\end{equation}\\
We remind that $\alpha_{\text{out}} (t_{\text{out}})=\rho_{\text{out}_1}e^{i\eta_{\text{out}_1}}$ and $\beta_{\text{out}} (t_{\text{out}})=\rho_{\text{out}_2}e^{i\eta_{\text{out}_2}}$  wherein
\begin{footnotesize}
\begin{align*}
 \rho_{\text{out}_1}&=\frac{\sqrt{2}}{2} \sqrt{N + \frac{R_{max}}{m}},\\
 \rho_{\text{out}_2}&=\frac{\sqrt{2}}{2} \sqrt{N - \frac{R_{max}}{m}}, \\
 \eta_{\text{out}_1}- \eta_{\text{out}_2}  &= \frac{\pi}{2} -\Phi_{out}(t_{\text{out}}),
\end{align*}
\end{footnotesize}
and
$$\Phi_{out}(t_{\text{out}})=2mt_{\text{out}} + \phi_{\text{max}} - 2mt_{\text{max}}$$
In the same way, the pre-selected state is given by:

$$\vec{w_i}= 
\begin{pmatrix}
	\langle \xi_{\text{in}} (t)| \sigma^1 | \xi_{\text{in}}(t) \rangle \\
	\langle \xi_{\text{in}}(t) | \sigma^2 | \xi_{\text{in}}(t) \rangle \\
	\langle \xi_{\text{in}}(t) | \sigma^3 | \xi_{\text{in}}(t) \rangle
\end{pmatrix}, 
$$
where $\xi_{\text{in}} (t) = \left( \begin{array}{c}
	\alpha_{\text{in}} (t)	\\
	\beta_{\text{in}}(t)
\end{array}\right)$ is expressed as

{\small \begin{equation}
		\begin{split}
			\alpha_{\text{in}} (t)=	&\frac{\sqrt{2}}{2}\alpha_{\text{in}} (t_{\text{in}}) \left[  \cos{(g(t))} e^{-i F(t, t_{\text{in}})}  +  \sin{( g(t))} e^{i F(t, t_{\text{in}})} \right]  +\\ &\frac{\sqrt{2}}{2}\beta_{\text{in}} (t_{\text{in}}) \left[  -\cos{( g(t))} e^{-i F(t, t_{\text{in}})}  +  \sin{(g(t))} e^{i F(t, t_{\text{in}})} \right],
		\end{split}
	\end{equation}}	
and 
{\small \begin{equation}
		\begin{split}
			\beta_{\text{in}} (t)=&	\frac{\sqrt{2}}{2}\alpha_{\text{in}} (t_{\text{in}}) \left[ - \sin{(g(t))} e^{-i F(t, t_{\text{in}})}  + \cos{( g(t))} e^{i F(t, t_{\text{in}})} \right]  +\\ &\frac{\sqrt{2}}{2}\beta_{\text{in}} (t_{\text{in}}) \left[  \sin{( g(t))} e^{-i F(t, t_{\text{in}})}  +  \cos{(g(t))} e^{i F(t, t_{\text{in}})} \right].
		\end{split}
\end{equation}}
 
 Let us remind that $\alpha_{\text{in}} (t_{\text{in}})=\rho_{\text{in}_1}e^{i\eta_{\text{in}_1}}$ and $\beta_{\text{in}} (t_{\text{in}})=\rho_{\text{in}_2}e^{i\eta_{\text{in}_2}}$  wherein
\begin{footnotesize}
\begin{align*}
 \rho_{\text{in}_1}&=\frac{\sqrt{2}}{2}\sqrt{N  + \sqrt{N^2 -\frac{R^4_0}{\lambda^2}}\cos{ \left( \Phi_{in}(t_{\text{in}})\right)}},\\
 \rho_{\text{in}_2}&=\frac{\sqrt{2}}{2}\sqrt{N  - \sqrt{N^2 -\frac{R^4_0}{\lambda^2}}\cos{ \left( \Phi_{in}(t_{\text{in}})\right)}}, \\
 \eta_{in_1}- \eta_{in_2}  &= -\arctan \left( \frac{ \sqrt{\lambda^2N^2 -R^4_0}\sin{\left(\Phi_{in}(t_{\text{in}})\right)}}{R^2_0}\right),
\end{align*}
\end{footnotesize}
and 
\begin{align*}
&\Phi_{in}(t_{\text{in}})=4 \lambda G(t_{\text{in}})-\lambda \pi -\phi_0, \\
& \text{and } G(t_{\text{in}})=  \tan^{-1}\left(\frac{t_{\text{in}}}{R(t_{\text{in}})}\right).
\end{align*}

 We obtain the following post-selected  $\vec{w_f} $  whose components are given as
\smallskip
	$\vec{w_f}=$
\begin{small}
\begin{eqnarray}
	\left\{
	\begin{array}{lll}
		&{w_f}^1 = -\frac{R_{\text{max}}}{m}  \sin{(2g(t))} + A \cos{\left(\Delta_{\text{out}} -2F(t_{\text{out}}, t)\right)}\cos{(2g(t))}, \\
		&{w_f}^2 =   A \sin{(\Delta_{\text{out}}  -2F(t_{\text{out}}, t))},  \\
	    &{w_f}^3 = \frac{R_{\text{max}}}{m} \cos{(2g(t))} + A\cos{(\Delta_{\text{out}}  -2F(t_{\text{out}}, t))}\sin{(2g(t))},\\
	\end{array}
	\right.,
\end{eqnarray}	
\end{small}
	wherein we have set for readability $A=\left(N^2- \frac{R^2_{max}}{m^2}\right)^{\frac{1}{2}} $
and $\Delta_{\text{out}}=\eta_{\text{out}_1}- \eta_{\text{out}_2}  $. 

We obtain the pre-selected vector  $\vec{w_i} $ whose components are given as 
\begin{small}
$\vec{w_i}=$
	\begin{eqnarray}
	\left\{
	\begin{array}{lll}
		w_i^1 &=&  \sqrt{N^2 -\frac{R_0^4}{\lambda^2}}\cos{(2g(t))} \cos{(2F(t, t_{\text{in}}))}+\frac{R_0^2}{\lambda} \sin{(2g(t))}  \\
		w_i^2 &=& -\sqrt{N^2 -\frac{R_0^4}{\lambda^2}} \sin{(2F(t, t_{\text{in}}))}  \\
		w_i^3 &=& \sqrt{N^2 -\frac{R_0^4}{\lambda^2}} \sin{(2g(t))} \cos{(2F(t, t_{\text{in}}))}  -\frac{R_0^2}{\lambda} \cos{(2g(t))} 
	\end{array}
	\right. .
\end{eqnarray}
		\end{small}

wherein we have considered the case $\Phi_{in}(t_{\text{in}})=0$. This is possible for $t_{in}=t_B $ (see (\ref{ScaleIn}) ) and $\phi_0=0 $.

The vector $\vec{f}$ is equal to $ \frac{\vec{w_f}}{|\vec{w_f}|}, \text{and } \vec{i}=\frac{\vec{w_i}}{|\vec{w_i}|}$. The length of both Bloch vectors are 
$$|\vec{w_f}|=|\vec{w_i}|=\lambda^2 - \frac{1}{4}.$$ They  represent the unit vectors on the Bloch sphere of post-selected  and pre-selected state. 
In terms of the components, the $\sigma^z$ weak measurement is given by

\begin{equation}
	\sigma _{r,w} =  \frac{{w_f}^3 +{w_i}^3 + i(-{w_f}^1{w_i}^2 + {w_f}^2{w_i}^1)}{\frac{1}{2} (1+ {w_f}^1{w_i}^1 + {w_f}^2{w_i}^2 +{w_f}^3{w_i}^3 )}.    \label{eq:weakmeas6}
\end{equation}

 The real part of the  numerator of the weak value of $\sigma^z$ operator 
in terms of the pre-selected and post-selected we have computed will be given as
${w_f}^3 +{w_i}^3=$
{\footnotesize 
	\begin{align}
	&\sin{(2g(t)}\left(\sqrt{N^2 -\frac{R_0^4}{\lambda^2}} \cos{(2F(t, t_{\text{in}}))}+  A \cos{\left(\Delta_{\text{out}} -2F(t_{\text{out}}, t)\right)}\right)\\ \nonumber
	& +\cos{(2g(t))}\left(\frac{R_{\text{max}}}{m}-  \frac{R_0^2}{\lambda}  \right).
	\end{align}}

 In the same way, the imaginary part of the numerator of the weak value of $\sigma^z$ operator  
in terms of the pre-selected and post-selected state  is given as\\
$-{w_f}^1{w_i}^2 + {w_f}^2{w_i}^1=$
{\footnotesize 
	\begin{equation}
	\begin{split}
	&\sin{ (2 g(t))} \times\\
		& \left(-\frac{R_{\text{max}}}{m} \sqrt{N^2-\frac{R_0^4}{\lambda^2}}\sin{ \left(2 F\left(t,t_{\text{in}}\right)\right)}  + A\frac{R_0^2}{\lambda}  \sin{(\Delta_{\text{out}}  -2F(t_{\text{out}}, t)} \right)\\
		& +  A\sqrt{N^2 -\frac{R_0^4}{\lambda^2}}\cos{(2g(t))} \sin{(\Delta_{\text{out}}+2 F\left(t,t_{\text{in}}\right) +2 F\left(t,t_{\text{out}}\right) )}.
	\end{split}
\end{equation}
}

Finally, the denominator part will be given in general terms by:\\
$ (1+ {w_f}^1{w_i}^1 + {w_f}^2{w_i}^2 +{w_f}^3{w_i}^3 )=$
%
	
	
	\begin{small}
	\begin{equation}
		\begin{split}
		& \sqrt{N^2-\frac{R_0^4}{\lambda ^2}} \sqrt{N^2-\frac{R_{\max }^2}{m^2}} \cos \left(a\right)-\frac{R_0^2 R_{\max }}{\lambda  m}+1
	\end{split}.
	\end{equation}
	\end{small}
wherein $a=2 F\left(t,t_{\text{in}}\right)+2 F\left(t,t_{\text{out}}\right)+\Delta _{\text{out}}$.

One can see that the amplification phenomenon is obtained in many ways. One way is 
	\begin{equation}
		\begin{split}
		\cos(a) \approx -\frac{1-\frac{R_0^2 R_{\max }}{\lambda  m}}{\sqrt{N^2-\frac{R_0^4}{\lambda ^2}} \sqrt{N^2-\frac{R_{\max }^2}{m^2}}} 
	\end{split}.
	\end{equation}
We remind that $\Delta _{\text{out}}=\eta_{out_1} - \eta_{out_2} $. By adjusting adequately the parameters appearing in $a$, one can amplify the weak values.

It is essential to note that any arbitrary observable in a 2-level system can be expressed as $\hat{A} = a_I \hat{I}_2 + a_L \vec{\alpha} .\hat{\vec{\sigma}}$, where $a_I$ and $a_L$ are constants, $\vec{\alpha}$ is a unit vector in three dimensions, and $\sigma$ denotes the Pauli matrices. 

In this subsection we have derived the computation of the weak value of the Z component of the Spin of fermions in the context of cosmology. We have shown how it is possible to amplify the signal by choosing adequate parameters.

\subsection{Berry phase from the weak value of a pure state. }

Let us now compute the weak value of a pure state which is given as: 
\begin{equation}
	\hat{\Pi }_{r,w} = \frac{\langle \psi_{out}(t)| \hat{\Pi}_r |\psi_{in}(t)\rangle}{\langle \psi_{out}(t) | \psi_{in}(t)\rangle}. \label{eq:weakmeas2}
\end{equation}

where $\hat{\Pi}_r $ is a pure state given as $ \frac{1}{2}\left(\hat{I }+\vec{r}\vec{\hat{\sigma}} \right) $. The complete formula to compute the weak values of a pure state is found in \cite{kofman2012nonperturbative} or the appendix of \cite{cormann2017geometric} and \cite{ferraz2022geometrical}. It reads as

\begin{equation}
	\hat{\Pi }_{r,w} = \frac{1+ \vec{f}. \vec{r} + \vec{r}. \vec{i} + \vec{f} .\vec{i} + i\vec{f}.(\vec{r} \wedge \vec{i})}{\frac{1}{2} (1 + \vec{f} .\vec{i} )}.    \label{eq:weakmeas3}
\end{equation}

The vector $\vec{r}$ is defined as 
\begin{equation}
	 \left( \begin{array}{c}
		\sin(\theta)\cos(\phi) \\
		\sin(\theta) \sin(\phi) \\
		\cos(\theta)
	\end{array}\right). 
\end{equation}


We obtain in terms of the pre-selected and post-selected state computed for  $\Phi_{in}(t_{\text{in}})=0$ and $\theta=\varphi=\frac{\pi}{2}$ 

\begin{small}
	\begin{equation}
		\begin{split}
	1 +A\sin{(\Delta_{\text{out}}  -2F(t_{\text{out}}, t)}  -\frac{R_0^2 R_{\max }}{\lambda  m}\\
	+\left( A\cos{(a)}- \sin{(2F(t_{\text{out}}, t))}\right)\sqrt{N^2-\frac{R_0^4}{\lambda ^2}}
	\end{split}.
	\end{equation}
	\end{small}

The imaginary part of numerator of the $\hat{\Pi }_{r,w} $ is given as

\begin{equation}
	\begin{split}
&-\frac{R_{\max }}{m}\sqrt{N^2-\frac{R_0^4}{\lambda ^2}} \cos \left(2 F\left(t,t_{\text{in}}\right)\right)\\
&-\frac{R_0^2 }{\lambda } \sqrt{N^2-\frac{R_{\max }^2}{m^2}}\cos{ \left(2F(t_{\text{out}}, t)-\Delta _{\text{out}}\right)}.
\end{split}
\end{equation}
The denominator, $1 + \vec{f} .\vec{i}$ remains the same as computed previously
%
	\begin{small}
	\begin{equation}
		\begin{split}
		& \sqrt{N^2-\frac{R_0^4}{\lambda ^2}} \sqrt{N^2-\frac{R_{\max }^2}{m^2}} \cos \left(a\right)-\frac{R_0^2 R_{\max }}{\lambda  m}+1
	\end{split}.
	\end{equation}
	\end{small}

The amplification phenomenon is obtained as explained in the case of the weak value of $\sigma^z$. 

Let us compute now the argument of the weak value of the pure state. It is given from \cite{ferraz2022geometrical} as

\begin{equation}
\arg	\hat{\Pi }_{r,w} = \arctan \frac{\vec{f}.(\vec{r} \wedge \vec{i})}{1+ \vec{f}. \vec{r} + \vec{r}. \vec{i} + \vec{f} .\vec{i} }    \label{eq:weakmeas4}
\end{equation}

The berry phase  would be given by $\arg	\hat{\Pi }_{r,w} =$
\begin{footnotesize}
\begin{equation}
	\arctan {\left( \frac{-\frac{R_{\max }}{m} \sqrt{N^2-\frac{R_0^4}{\lambda ^2}} \cos \left(2 F\left(t,t_{\text{in}}\right)\right)-\frac{R_0^2}{\lambda } A\cos \left(b\right)}{1 -A\sin{(b)}  -\frac{R_0^2 R_{\max }}{\lambda  m} +\left( A\cos{(a)}- \sin{(2F(t_{\text{out}}, t))}\right)\sqrt{N^2-\frac{R_0^4}{\lambda ^2}}}  \right)}. 
\end{equation}
\end{footnotesize}
wherein $ b=2F(t_{\text{out}}, t)-\Delta _{\text{out}}$.

As a conclusion, this berry phase confirms the fact that the Universe considered here is not flat since its argument is not 0 or a multiple of $2\pi$ in all cases	.

\section{CONCLUSION}

Our objectives in this paper were to apply the theory of weak measurement in Einstein-Dirac system in the \emph{Friedman-Lemaître-Robertson-Walker} space. We begun by noticing with the help of Davies \cite{davies2014quantum} that all observations and measurement are weak in the quantum sense due to the nature of the processes involve. Then, after computing the two asymptotic state vector, solutions of the Einstein-Dirac equations around the Big Bang and the Universe at its maximal value of the scale factor and finding the evolutionary operator, we computed the weak values of the energy momentum tensor and two other operators.
 
In this work, we have elucidated numerous conclusions, particularly regarding the enlarged scope of weak measurements in contrast to classical measurement paradigms. Notably, our analysis revealed that the weak value of the energy-momentum tensor extends beyond what is described classical energy-momentum concepts, exhibiting unusual values as complex numbers in certain instances. 

We have also shown that it is possible to amplify  measurements with less expensive equipments through the weak measurements process by strategically assuming appropriate initial and final state vector. Furthermore, weak measurements serve as a powerful tool for shedding some lights to the geometric characteristics of the underlying space. By virtue of its association with the Berry phase, weak measurements enabled us to ascertain that the underlying space exhibits non-flat geometry, as evidenced by the weak measurement of fermion wave functions.

Our analytical framework also demonstrates the effectiveness of weak measurements in  detecting the acceleration of the Universe, leveraging the Two-State vector formalism and weak measurement techniques. Notably, we established that the scale function may be understood as an outcome of the post-selection process.  

We end this paper by noticing that it is possible to use this tool experimentally in cosmology. This should be the next step for some future researches.
	
	\begin{acknowledgements}
	We gratefully acknowledge the Fund for Scientific Research (FNRS) grant and the Society of Jesus to have funded our research stay at the University of Regensburg to finalize this article. We extend our deep gratitude to Professor Felix Finster for his invaluable insights and engaging discussions during the finalization of this work. We thank the Department of Mathematics at the University of Regensburg for providing a conducive environment during our research. Additionally, we thank Ramachandran Chittur Anantharaman and Charles Modera for checking the orthographies and grammar of this paper.
	\end{acknowledgements}
	
	\appendix
\begin{widetext}
	\section{ Weak measurements of the Energy momentum  }

Let us make the computations of the energy momentum more explicit where $ \langle  T_{0}^{0} \rangle_w = \frac{\langle \psi_{out}(t)| T_{0}^{0} |\psi_{in}(t)\rangle }{\langle \psi_{out}(t) | \psi_{in}(t)\rangle}$ and 
	$ \langle  T_{j}^{j} \rangle_w =	\frac{\langle \psi_{out}(t)|  T_{j}^{j}|\psi_{in}(t)\rangle }{\langle \psi_{out}(t) | \psi_{in}(t)\rangle}.$ They are given as follows
	
			\begin{equation}
				\begin{aligned}
					& \langle  T_{0}^{0} \rangle_w =
					\frac{\Bigl[ m R(t) \left(\alpha_{\text{in}}(t) \overline{\alpha_{\text{out}}}(t) - \beta_{\text{in}} (t)\overline{\beta_{\text{out}}}(t)\right) -\lambda \left(   \alpha_{\text{in}(t)} \overline{\beta_{\text{out}}}(t) +  \beta_{\text{in}}(t) \overline{\alpha_{\text{out}}}(t)  \right) \Bigr] }{R(t)\left(\overline{\alpha_{\text{out}}}(t) \alpha_{\text{in}}(t) + \overline{\beta_{\text{out}}}(t) \beta_{\text{in}}(t) \right) } \\
				\end{aligned},
			\end{equation}

		and 
	
			\begin{equation}
				\begin{aligned}
					&\langle  T_{j}^{j} \rangle_w =  
					\frac{\lambda \left(   \alpha_{\text{in}}(t) \overline{\beta_{\text{out}}}(t) +  \beta_{\text{in}}(t) \overline{\alpha_{\text{out}}}(t)  \right) }{3R(t)\left(\overline{\alpha_{\text{out}}}(t) \alpha_{\text{in}}(t) + \overline{\beta_{\text{out}}}(t)\beta_{\text{in}} (t)\right)}\\
				\end{aligned}.
			\end{equation}
	\end{widetext}
		\subsection{Computation of the pre-selected state at time t}

The spinor at time $t$ is obtained by applying the evolutionary operator on a spinor defined at a specific time. The case of pre-selected state at time $t$ is the pre-selected  state at time $t_{\text{in}}$ on which the unitary operator is applied. It is given by
$$\left.  \left( \begin{array}{c}
	\alpha_{\text{in}} (t)	\\
	\beta_{\text{in}}(t)
\end{array}\right)=U(t, t_{\text{in}} ) \left( \begin{array}{c}
	\alpha_{\text{in}} (t_{\text{in}})	\\
	\beta_{\text{in}}(t_{\text{in}})
\end{array}\right) \right. $$.
\newline
$U(t, t_{\text{in}} ) $ is given by $U^{-1}(t)\left( \begin{array}{cc}
	e^{-iF(t ,t_{\text{in}} )} & 0 \\
	0 & e^{iF(t,t_{\text{in}} )}
\end{array}\right)U(t_{in})$ 


The specific case wherein the pre-selected state is taken in the radiation dominated Universe with $m=0$ will yield $g(t_{\text{in}})=\frac{1}{2} \arctan(\frac{\lambda}{0* R(t_{\text{in}})})= \frac{\pi}{4} + k \pi$. For $k=0$, the matrix will be given by $
U(t_{\text{in}})= \frac{\sqrt{2}}{2}\begin{pmatrix}
	1 & -1 \\
	1 & 1
\end{pmatrix}  $.

This will lead to\\
{\small \begin{equation}
		\begin{split}
			\alpha_{\text{in}} (t)=	&\frac{\sqrt{2}}{2}\alpha_{\text{in}} (t_{\text{in}}) \left[  \cos{(g(t))} e^{-i F(t, t_{\text{in}})}  +  \sin{( g(t))} e^{i F(t, t_{\text{in}})} \right]  +\\ &\frac{\sqrt{2}}{2}\beta_{\text{in}} (t_{\text{in}}) \left[  -\cos{( g(t))} e^{-i F(t, t_{\text{in}})}  +  \sin{(g(t))} e^{i F(t, t_{\text{in}})} \right]  .
		\end{split}
	\end{equation} }	
and 
{\small \begin{equation}
		\begin{split}
			\beta_{\text{in}} (t)=&	\frac{\sqrt{2}}{2}\alpha_{\text{in}} (t_{\text{in}}) \left[ - \sin{(g(t))} e^{-i F(t, t_{\text{in}})}  + \cos{( g(t))} e^{i F(t, t_{\text{in}})} \right]  +\\ &\frac{\sqrt{2}}{2}\beta_{\text{in}} (t_{\text{in}}) \left[  \sin{( g(t))} e^{-i F(t, t_{\text{in}})}  +  \cos{(g(t))} e^{i F(t, t_{\text{in}})} \right]  .
		\end{split}
\end{equation} }

We remind here that $F(t, t_{\text{in}})=\int ^t _{t_{\text{in}}} \frac{\sqrt{R^2 (t)m^2 + \lambda^2}}{R(t)}dt$.

	\subsection{Computation of the post-selected state at time t}
The post-selected state is taken from the dust dominated Universe wherein $\lambda=0$. This will yield    $g(t)=\frac{1}{2} \arctan(\frac{0}{m* R(t_{\text{in}})})=0 + k\pi$ . This will lead,  for $k=0$, to $U(t_{out})= \left( \begin{array}{cc}
	1 & 0 \\
	0 & 1
\end{array}\right)$,  the identity matrix.

Knowing that $\left. U^{\dagger}(t, t_{\text{out}})=U(t_{\text{out}}, t)\right.  $, 
$$\left.  \left( \begin{array}{cc}
	\overline{	\alpha_{\text{out}}} (t)& \overline{\beta_{\text{out}}}(t)\end{array}\right) \right.=\left( \begin{array}{cc}
\overline{	\alpha_{\text{out}}} (t_{\text{out}})& \overline{\beta_{\text{out}}}(t_{\text{out}})
\end{array}\right) U(t_{\text{out}}, t).$$
\newline
$U(t_{\text{out}}, t) $ is given by $U^{-1}(t_{\text{out}})\left( \begin{array}{cc}
	e^{-iF(t _{\text{out}},t)} & 0 \\
	0 & e^{iF(t_{\text{out}},t)}
\end{array}\right)U(t).$ 

This will lead to:\\
{\footnotesize \begin{equation}
		\begin{split}
		\overline{	\alpha_{\text{out}}} (t)=\overline{\alpha_{\text{out}}} (t_{\text{out}})  \cos{(g(t))} e^{-i F(t_{\text{out}}, t)}  + \overline{ \beta_{\text{out}}} (t_{\text{out}})\sin{( g(t))} e^{i F(t_{\text{out}}, t)}   .
		\end{split}
\end{equation} }
and :\\
{\footnotesize \begin{equation}
		\begin{split}
			 \overline{\beta_{\text{out}}}(t)=-\overline{\alpha_{\text{out}}} (t_{\text{out}})  \sin{(g(t))} e^{-i F(t_{\text{out}}, t)}  + \overline{ \beta_{\text{out}}} (t_{\text{out}})\cos{( g(t))} e^{i F(t_{\text{out}}, t)}   .
		\end{split}
\end{equation} }

	\subsection{ Computation of  numerators and denominators of the energy-momentum Tensor }
		In terms of $\alpha_{\text{in}}(t_{\text{in}}) \quad \text{and} \quad \alpha_{\text{out}}(t_{\text{out}})$ \\where
\begin{align*}
    A &= e^{i\left(F(t_{\text{out}}, t_{\text{in}}) \right)},  \quad \bar{A}= e^{-i\left(F(t_{\text{out}}, t_{\text{in}}) \right)}, \\
    B &= e^{i\left( F(t, t_{\text{in}}) + F(t, t_{\text{out}}) \right)}, \quad \bar{B} = e^{-i\left( F(t, t_{\text{in}}) + F(t, t_{\text{out}}) \right)}
\end{align*}

		\begin{itemize}
			\item 	\textbf{$\left(\overline{\alpha_{\text{out}}}(t) \alpha_{\text{in}}(t) + \overline{\beta_{\text{out}}}(t) \beta_{\text{in}}(t)\right)$ }is equal to:
			{\footnotesize 
				\begin{equation}
					\begin{split}
						&	\frac{\sqrt{2}}{2}\left(\overline{\alpha_{\text{out}}}(t_{\text{out}}) \alpha_{\text{in}}(t_{\text{in}}) - \overline{\alpha_{\text{out}}}(t_{\text{out}}) \beta_{\text{in}}(t_{\text{in}})\right) \bar{A}  \\
						&+	\frac{\sqrt{2}}{2}\left(\overline{\beta_{\text{out}}}(t_{\text{out}}) \beta_{\text{in}}(t_{\text{in}}) + \overline{\beta_{\text{out}}}(t_{\text{out}}) \alpha_{\text{in}}(t_{\text{in}})\right) A.
					\end{split}
				\end{equation}}
			\item \textbf{ $\left(\alpha_{\text{in}}(t) \overline{\alpha_{\text{out}}}(t) - \beta_{\text{in}} (t)\overline{\beta_{\text{out}}}(t)\right)$ }is equal to:
			{\footnotesize\begin{equation}
					\begin{split}
						&\frac{\sqrt{2}}{2} \alpha_{\text{in}}(t_{\text{in}}) \overline{\alpha_{\text{out}}}(t_{\text{out}}) \left[\sin{(2 g(t))} B + \cos{(2 g(t))} \bar{A} \right] \\
						&+ \frac{\sqrt{2}}{2}\beta_{\text{in}}(t_{\text{in}}) \overline{\beta_{\text{out}}}(t_{\text{out}}) \left[-\sin{(2 g(t))} \bar{B}- \cos{(2 g(t))} A\right] \\
						&+ \frac{\sqrt{2}}{2}\alpha_{\text{in}}(t_{\text{in}}) \overline{\beta_{\text{out}}} (t_{\text{out}})\left[\sin{(2 g(t))} \bar{B} - \cos{(2 g(t))} A\right] \\
						&+ \frac{\sqrt{2}}{2} \beta_{\text{in}}(t_{\text{in}}) \overline{\alpha_{\text{out}}} (t_{\text{out}})\left[\sin{(2 g(t))} B - \cos{(2 g(t))} \bar{A} \right].
					\end{split}
				\end{equation}}
			\item \textbf{$\left(\alpha_{\text{in}}(t) \overline{\beta_{\text{out}}}(t) + \beta_{\text{in}} (t)\overline{\alpha_{\text{out}}}(t)\right)$ }is equal to:
			{\footnotesize
				\begin{equation}
					\begin{split}
						& \frac{\sqrt{2}}{2}\alpha_{\text{in}}(t_{\text{in}}) \overline{\alpha_{\text{out}}}(t_{\text{out}}) \left[- \sin{(2 g(t))} \bar{A}+\cos{(2 g(t))}B\right] \\
						&+ \frac{\sqrt{2}}{2}\beta_{\text{in}}(t_{\text{in}}) \overline{\beta_{\text{out}}}(t_{\text{out}}) \left[ \sin{(2 g(t))} A-\cos{(2 g(t))} \bar{B} \right] \\
						&+ \frac{\sqrt{2}}{2}\alpha_{\text{in}}(t_{\text{in}}) \overline{\beta_{\text{out}}} (t_{\text{out}})\left[\sin{(2 g(t))} A+ \cos{(2 g(t))}\bar{B}  \right] \\
						&+\frac{\sqrt{2}}{2} \beta_{\text{in}}(t_{\text{in}}) \overline{\alpha_{\text{out}}} (t_{\text{out}})\left[\sin{(2 g(t))} \bar{A} + \cos{(2 g(t))} B \right].
					\end{split}
				\end{equation}	}
	\end{itemize}
	

	\section{Weak measurements of $\sigma^z$ operator }
	
The weak measurement of the generator of $SU(2)$ is given by 
	
	\begin{equation}
		\sigma _{r,w} =  \frac{ \vec{f}. \vec{r} + \vec{r}. \vec{i} + i\vec{f}.(\vec{r} \wedge \vec{i})}{\frac{1}{2} (1 + \vec{f} .\vec{i} )}.    \label{eq:weakmeas6}
	\end{equation}

	In our case, $\vec{r}=(\begin{array}{ccc}	0 & 0 & 1\end{array})^T$ to get $\sigma^z$.  Following Kofman  in  \cite{kofman2012nonperturbative}, the vectors $ \vec{f}$ and $ \vec{i}$ are computed as follow:
	$\vec{w_f}=$
	$$ 
	\begin{pmatrix}
		\langle \xi_{\text{out}} (t)| \sigma^1 | \xi_{\text{out}}(t) \rangle \\
		\langle \xi_{\text{out}}(t) | \sigma^2 | \xi_{\text{out}}(t) \rangle \\
		\langle \xi_{\text{out}}(t) | \sigma^3 | \xi_{\text{out}}(t) \rangle
	\end{pmatrix} =
	\left( \begin{array}{c} 
		2\text{Re}\left(\alpha_{\text{out}}(t)\bar{\beta}_{\text{out}}(t)\right)  
		\\	2\text{Im}\left(\alpha_{\text{out}}(t)\bar{\beta}_{\text{out}}(t)\right)
		\\ |\alpha_{\text{out}(t)}|^2 - |\beta_{\text{out}(t)}|^2
	\end{array}\right). 
	$$
	wherein $\xi_{\text{out}} (t) = \left( \begin{array}{c}
		\alpha_{\text{out}} (t)	\\
		\beta_{\text{out}}(t)
	\end{array}\right).$
	
	Some particular products are given as:\\

\begin{small}
\begin{equation}
\begin{split}
		&\alpha_{\text{out}}(t)\bar{\beta}_{\text{out}}(t)= -\left( |\alpha_{\text{out}}(t_{\text{out}})|^2 -  |\beta_{\text{out}}(t_{\text{out}})|^2 \right) \sin(g(t))  \cos(g(t)) \\
		&+\overline{ \alpha_{\text{out}}}(t_{\text{out}})\beta_{\text{out}}(t_{\text{out}}) \cos^2(g(t))e^{-2 iF(t_{\text{out}}, t)}\\
		& -\overline{ \beta_{\text{out}}}(t_{\text{out}})\alpha_{\text{out}}(t_{\text{out}}) \sin^2(g(t))e^{2 iF(t_{\text{out}}, t)},
\end{split}.
\end{equation}

\end{small}
In terms of $\alpha_{\text{out}}(t_{\text{out}})= \rho_ {out_1}e^{i\eta_{out_1}}$ and $\beta_{\text{out}}(t_{\text{out}})= \rho_ {out_2}e^{i\eta_{out_2}}$, the real part  will be 

{\small 
	\begin{equation}
		\begin{split}
			&\text{Re}\left(\alpha_{\text{out}}(t)\bar{\beta}_{\text{out}}(t)\right) =  \frac{1}{2}\left(  \rho_{out_2}^2 -\rho_{out_1}^2\right) \sin{(2g(t))} \\
			& + \rho_{out_1} \rho_{out_2} \cos{(2 g(t))} \cos {(\eta_{out_2} - \eta_{out_1} - 2 F(t_{\text{out}}, t))},
		\end{split}
\end{equation}}

and the imaginary part in terms of $\rho_{out_1},  \rho_{out_2}, \eta_{out_2}, \eta_{out_1} $ will be
	
	{\small
		\begin{equation}
			\begin{split}
				&\text{Im}\left(\alpha_{\text{out}}(t)\bar{\beta}_{\text{out}}(t)\right) =  \frac{1}{2}\left(  \rho_{out_2}^2 -\rho_{out_1}^2\right) \sin{(2g(t))} \\
				& + \rho_{out_1} \rho_{out_2} \sin{(\eta_{out_2} - \eta_{out_1} - 2 F(t_{\text{out}}, t))}.
			\end{split}
	\end{equation}}

	In the same way, the pre-selected state vector is computed as follows 
	$\vec{w_i}=$
	$$ 
	\begin{pmatrix}
		\langle \xi_{\text{in}} (t)| \sigma^1 | \xi_{\text{in}}(t) \rangle \\
		\langle \xi_{\text{in}}(t) | \sigma^2 | \xi_{\text{in}}(t) \rangle \\
		\langle \xi_{\text{in}}(t) | \sigma^3 | \xi_{\text{in}}(t) \rangle
	\end{pmatrix} =
	\left( \begin{array}{c} 
		2\text{Re}\left(\alpha_{\text{in}}(t)\bar{\beta}_{\text{in}}(t)\right)  
		\\	2\text{Im}\left(\alpha_{\text{in}}(t)\bar{\beta}_{\text{in}}(t)\right)
		\\ |\alpha_{\text{in}(t)}|^2 - |\beta_{\text{in}(t)}|^2
	\end{array}\right), 
	$$
	wherein $\xi_{\text{in}} (t) = \left( \begin{array}{c}
		\alpha_{\text{in}} (t)	\\
		\beta_{\text{in}}(t)
	\end{array}\right)$
	
{\footnotesize	\begin{equation}
		\begin{split}
			&\alpha_{\text{in}}(t)\bar{\beta}_{\text{in}}(t) \\
			&= \left( |\alpha_{\text{in}}(t_{\text{in}})|^2 - |\beta_{\text{in}}(t_{\text{in}})|^2 \right) \\
			&\quad \times \Biggl( \cos^2(g(t))e^{-2iF(t, t_{\text{in}})} - \sin^2(g(t))e^{2iF(t, t_{\text{in}})} \Biggr) \\
			&+ \alpha_{\text{in}}(t_{\text{in}})\overline{\beta}_{\text{in}}(t_{\text{in}}) \\
			&\quad \times \Biggl( \sin(2g(t)) + \sin^2(g(t))e^{2iF(t, t_{\text{in}})} + \cos^2(g(t))e^{-2iF(t, t_{\text{in}})} \Biggr) \\
			&+ \beta_{\text{in}}(t_{\text{in}})\overline{\alpha}_{\text{in}}(t_{\text{in}}) \\
			&\quad \times \Biggl( \sin(2g(t)) - \sin^2(g(t))e^{2iF(t, t_{\text{in}})} - \cos^2(g(t))e^{-2iF(t, t_{\text{in}})} \Biggr),
		\end{split}.
	\end{equation}
}
	In terms of $\alpha_{\text{in}}(t_{\text{in}})= \rho_ {in_1}e^{\eta_{in_1}}$ and $\beta_{\text{in}}(t_{\text{in}})= \rho_ {in_2}e^{\eta_{in_2}}$, the real part  will be 
	{\small
		\begin{equation}
			\begin{split}
				&\text{Re}\left(\alpha_{\text{in}}(t)\bar{\beta}_{\text{in}}(t)\right) =  \left( \rho_{in_1}^2 - \rho_{in_2}^2 \right) \cos{(2g(t))}\cos{(2F(t, t_{\text{in}}))} \\
				& + 2\rho_{in_1} \rho_{in_2} \left[ \sin{(2 g(t))} \cos {(\eta_{in_1} - \eta_{in_2})} \right. \\
				& \left. - \cos{(2 g(t))} \sin {(\eta_{in_1} - \eta_{in_2})} \sin{(2F(t, t_{\text{in}}))} \right] 
			\end{split}
		\end{equation}
		}

and the imaginary part will be 
		{\small 
		\begin{equation}
			\begin{split}
				&\text{Im}\left(\alpha_{\text{in}}(t)\bar{\beta}_{\text{in}}(t)\right) =  \left(  \rho_{in_2}^2 -\rho_{in_1}^2\right) \sin{(2F(t, t_{\text{in}}))}\\
				& + 2\rho_{in_1} \rho_{in_2}  \sin{(\eta_{in_1} - \eta_{in_2} )} \cos{(2F(t, t_{\text{in}}))}. 
			\end{split}
	\end{equation}}
	
Following Kofman  in  \cite{kofman2012nonperturbative}, the vectors $ \vec{f}$ and $ \vec{i}$ are computed as follows

$$\vec{w_f}= 
\begin{pmatrix}
	\langle \xi_{\text{out}} (t)| \sigma^1 | \xi_{\text{out}}(t) \rangle \\
	\langle \xi_{\text{out}}(t) | \sigma^2 | \xi_{\text{out}}(t) \rangle \\
	\langle \xi_{\text{out}}(t) | \sigma^3 | \xi_{\text{out}}(t) \rangle
	\end{pmatrix} =
	\left( \begin{array}{c} 
	2\text{Re}\left(\alpha_{\text{out}}(t)\bar{\beta}_{\text{out}}(t)\right)  
	\\	2\text{Im}\left(\alpha_{\text{out}}(t)\bar{\beta}_{\text{out}}(t)\right)
	\\ |\alpha_{\text{out}(t)}|^2 - |\beta_{\text{out}(t)}|^2
	\end{array}\right), 
$$
wherein $\xi_{\text{out}} (t) = \left( \begin{array}{c}
	\alpha_{\text{out}} (t)	\\
	\beta_{\text{out}}(t)
\end{array}\right)$. 
We obtain the following vector $\vec{w_f} $  components that are given as:

{\footnotesize \begin{eqnarray}
	\left\{
	\begin{array}{lll}
		{w_f}^1 &=&-\Delta \rho^2_{out} \sin{(2g(t))} 
		 + 2 \rho_{out_1} \rho_{out_2} \cos{(2 g(t))} \cos{(\Delta_{out} )} \\
		{w_f}^2 &=&  2 \rho_{out_1} \rho_{out_2} \sin{(\Delta_{out} )}\\
		{w_f}^3 &=& \Delta \rho^2_{out} \cos{(2g(t))}+ 2 \rho_{out_1} \rho_{out_2} \sin{(2 g(t))} \cos {(\Delta_{out} )},
	\end{array}
	\right.
\end{eqnarray}
}

where $(\eta_{out_1} - \eta_{out_2} +2 F(t_{\text{out}}, t))=\Delta_{out}$, \\$ \rho_{out_1}^2 -\rho_{out_2}^2=\Delta \rho^2_{out}$.\\
Its norm is $|\vec{w_f}| =\rho_{out_1}^2 +\rho_{out_2}^2 =\lambda^2 - \frac{1}{4}$. This implies that the unit vector $\vec{f}$ would be given by
$$ \vec{f}	=\frac{1}{\lambda^2 - \frac{1}{4}}\vec{w_f}.$$

%

In the same way, the pre-selected state is given by:

$$\vec{w_i}= 
\begin{pmatrix}
	\langle \xi_{\text{in}} (t)| \sigma^1 | \xi_{\text{in}}(t) \rangle \\
	\langle \xi_{\text{in}}(t) | \sigma^2 | \xi_{\text{in}}(t) \rangle \\
	\langle \xi_{\text{in}}(t) | \sigma^3 | \xi_{\text{in}}(t) \rangle
\end{pmatrix} =
\left( \begin{array}{c} 
	2\text{Re}\left(\alpha_{\text{in}}(t)\bar{\beta}_{\text{in}}(t)\right)  
	\\	2\text{Im}\left(\alpha_{\text{in}}(t)\bar{\beta}_{\text{in}}(t)\right)
	\\ |\alpha_{\text{in}(t)}|^2 - |\beta_{\text{in}(t)}|^2
\end{array}\right).
$$
wherein $\xi_{\text{in}} (t) = \left( \begin{array}{c}
	\alpha_{\text{in}} (t)	\\
	\beta_{\text{in}}(t)
\end{array}\right)$.\\
We obtain the following vector $\vec{w_{i}}= $ 

\begin{widetext}
	\begin{eqnarray}
		\left\{
		\begin{array}{lll}
			{w_i}^1 &=&  \Delta \rho_{in}^2  \cos{(2g(t))} \cos{(2F(t, t_{\text{in}}))}+ 2 \rho_{in_1} \rho_{in_2}   \left( \sin{(2g(t))} \cos{(\Delta_{in})} +\cos(2g)\sin(\Delta_{in}) \sin{(2F(t, t_{\text{in}}))} \right) \\
			{w_i}^2 &=& -\Delta \rho_{in}^2 \sin{(2F(t, t_{\text{in}}))}  + 2\rho_{in_1} \rho_{in_2}\sin{(\Delta_{in})} \cos{(2F(t, t_{\text{in}}))} \\
			{w_i}^3 &=& \Delta \rho_{in}^2 \sin{(2g(t))} \cos{(2F(t, t_{\text{in}}))}+ 2 \rho_{in_1} \rho_{in_2} \left(-\cos{(2g(t))} \cos{(\Delta_{in})} +\sin{(2g(t))} \sin{(\Delta_{in} )}  \sin{(2F(t, t_{\text{in}}))} \right),
		\end{array}
 		\right.
	\end{eqnarray}
\end{widetext}

wherein $\eta_{in_1} - \eta_{in_2}=\Delta_{in}$,  $ \rho_{in_1}^2 -\rho_{in_2}^2=\Delta \rho^2_{in}$,\\ $|\vec{w_f}|=\lambda^2 - \frac{1}{4}$. The vector $\vec{f}$ is equal to $ \frac{\vec{w_f}}{|\vec{w_f}|}, \text{and } \vec{i}=\frac{\vec{w_i}}{|\vec{w_i}|}$. They  represent the unit vectors on the Bloch sphere of post-selected  and pre-selected state and $\vec{r}=(\begin{array}{ccc}	0 & 0 & 1\end{array})$.  
In terms of the components, the $\sigma^z$ weak value is given by:
\begin{equation}
	\sigma _{r,w} =  \frac{{w_f}^3 +{w_i}^3 + i(-{w_f}^1{w_i}^2 + {w_f}^2{w_i}^1)}{\frac{1}{2} (1+ {w_f}^1{w_i}^1 + {w_f}^2{w_i}^2 +{w_f}^3{w_i}^3 )}    \label{eq:weakmeas6}
\end{equation}	
The  real and the imaginary part of the numerator of the weak measurement of $\sigma^z$ operator ,  $\sigma _{r,w}$, are  given respectively by: 	
\begin{widetext}
${w_f}^3 +{w_i}^3 =$	
	\begin{equation}
		\begin{split}
			& \Delta \rho _{\text{in}}^2 \cos \left(2 F\left(t,t_{\text{in}}\right)\right) \sin (2 g(t))+\Delta \rho _{\text{out}}^2 \cos (2 g(t))+2 \rho _{i n_1} \rho _{i n_2} \left(-\cos (2 g(t)) \cos \left(\Delta _{\text{in}}\right)+\sin \left(2 F\left(t,t_{\text{in}}\right)\right) \sin (2 g(t)) \sin \left(\Delta _{\text{in}}\right)\right)+\\
			&2 \rho _{o u t_1} \rho _{o u t_2} \sin (2 g(t)) \cos \left(\Delta _{\text{out}}\right),
		\end{split}
	\end{equation}
	$-{w_f}^1{w_i}^2 + {w_f}^2{w_i}^1=$
	\begin{footnotesize}
	\begin{equation}
		\begin{split}
			& 2 \rho _{i n_1} \rho _{i n_2} \left(\Delta \rho _{\text{out}}^2 \cos \left(2 F\left(t,t_{\text{in}}\right)\right) \sin (2 g(t)) \sin \left(\Delta _{\text{in}}\right)+2 \rho _{o u t_1} \rho _{o u t_2} \left(-\cos (2 g(t)) \cos \left(2 F\left(t,t_{\text{in}}\right)+\Delta _{\text{out}}\right) \sin \left(\Delta _{\text{in}}\right)+\cos \left(\Delta _{\text{in}}\right) \sin (2 g(t)) \sin \left(\Delta _{\text{out}}\right)\right)\right)+\\
			&\Delta \rho _{\text{in}}^2 \left(-\Delta \rho _{\text{out}}^2 \sin \left(2 F\left(t,t_{\text{in}}\right)\right) \sin (2 g(t))+2 \rho _{o u t_1} \rho _{o u t_2} \cos (2 g(t)) \sin \left(2 F\left(t,t_{\text{in}}\right)+\Delta _{\text{out}}\right)\right)
		\end{split},
	\end{equation}
	\end{footnotesize}
and $1+ {w_f}^1{w_i}^1 + {w_f}^2{w_i}^2 +{w_f}^3{w_i}^3 $ is given by:
\begin{equation}
		\begin{split}
1-2 \Delta \rho _{\text{out}}^2 \rho _{i n_1} \rho _{i n_2} \cos \left(\Delta _{\text{in}}\right)+2 \rho _{o u t_1} \rho _{o u t_2} \left(\Delta \rho _{\text{in}}^2 \cos \left(2 F\left(t,t_{\text{in}}\right)+\Delta _{\text{out}}\right)+2 \rho _{i n_1} \rho _{i n_2} \sin \left(\Delta _{\text{in}}\right) \sin \left(2 F\left(t,t_{\text{in}}\right)+\Delta _{\text{out}}\right)\right).
	\end{split}
	\end{equation}
\end{widetext}

\section{Berry phase }

The weak value of a pure is given by:
\begin{equation}
	\hat{\Pi }_{r,w} = \frac{1+ \vec{f}. \vec{r} + \vec{r}. \vec{i} + \vec{f} .\vec{i} + i\vec{f}.(\vec{r} \wedge \vec{i})}{\frac{1}{2} (1 + \vec{f} .\vec{i} )}.    
\end{equation}
We obtain from $ \vec{f}=\frac{\vec{w_f}}{|\vec{w_f}|}, \text{and } \vec{i}=\frac{\vec{w_i}}{|\vec{w_i}|}$ already computed in general terms:

\begin{widetext}
$1+ \vec{f}. \vec{r} + \vec{r}. \vec{i} + \vec{f} .\vec{i} =$	
\begin{equation}
		\begin{split}
		& 1+\cos (\varphi )\sin (\theta )\left(\Delta \rho _{\text{in}}^2 \cos (2 g(t)) \cos \left(2 F\left(t,t_{\text{in}}\right)\right)-\Delta \rho _{\text{out}}^2 \sin (2 g(t))\right.\\
		&	\left.+2 \rho _{i n_1} \rho _{i n_2} \left(\cos (2 g(t)) \sin \left(\Delta _{\text{in}}\right) \sin \left(2 F\left(t,t_{\text{in}}\right)\right)+\sin (2 g(t)) \cos \left(\Delta _{\text{in}}\right)\right)+2 \cos (2 g(t)) \cos \left(\Delta _{\text{out}}\right) \rho _{out_1} \rho _{out_2}\right)\\
		&+\cos (\theta )\left(\Delta \rho _{\text{in}}^2 \sin (2 g(t)) \cos \left(2 F\left(t,t_{\text{in}}\right)\right)+\Delta \rho _{\text{out}}^2 \cos (2 g(t))+2\rho _{i n_1}\rho _{i n_2}\left(-\cos (2 g(t)) \cos \left(\Delta _{\text{in}}\right)\right.\right.\\
		&\left.\left.\sin (2 g(t)) \sin \left(\Delta _{\text{in}}\right) +\sin \left(2 F\left(t,t_{\text{in}}\right)\right)\right)+2 \sin (2 g(t)) \cos \left(\Delta _{\text{out}}\right) \rho _{out_1} \rho _{out_2}\right)+\\
		&\sin (\theta ) \sin (\varphi ) \left(2 \rho _{i n_1} \rho _{i n_2} \sin \left(\Delta _{\text{in}}\right) \cos \left(2 F\left(t,t_{\text{in}}\right)\right)+\left(-\Delta \rho _{\text{in}}^2\right) \sin \left(2 F\left(t,t_{\text{in}}\right)\right) + 2 \sin \left(\Delta _{\text{out}}\right) \rho _{out_1} \rho _{out_2}\right)
\end{split}.
	\end{equation}
	
	The imaginary part, $\vec{f}.(\vec{r} \wedge \vec{i})$ is given by
	\begin{equation}
	\begin{split}
	&\sin (\theta ) \sin (\varphi ) \left(\left(-\Delta \rho _{\text{in}}^2\right) \Delta \rho _{\text{out}}^2 \cos \left(2 F\left(t,t_{\text{in}}\right)\right)-2 \rho _{i n_1} \rho _{i n_2} \left(\Delta \rho _{\text{out}}^2 \sin \left(\Delta _{\text{in}}\right) \sin \left(2 F\left(t,t_{\text{in}}\right)\right)+2 \cos \left(\Delta _{\text{in}}\right) \cos \left(\Delta _{\text{out}}\right) \rho_{out_1} \rho_{out_2 }\right)\right)\\
	& + \cos (\theta )\left(2\rho _{in_1}\rho _{i n_2}\Delta \rho _{\text{out}}^2\left(\sin (2 g(t)) \sin \left(\Delta _{\text{in}}\right) \cos \left(2 F\left(t,t_{\text{in}}\right)\right)+2\rho _{out_1}\rho _{out_2}\left(-\cos (2 g(t)) \sin \left(\Delta _{\text{in}}\right) \cos \left(2 F\left(t,t_{\text{in}}\right)+\Delta _{\text{out}}\right)\right.\right.\right.\\
	& +\left.\left.\left.\sin (2 g(t))\cos \left(\Delta _{\text{in}}\right) \sin \left(\Delta _{\text{out}}\right)\right)\right)+\Delta \rho _{\text{in}}^2 \left(2 \rho _{out_1} \rho _{out_2}\cos (2 g(t)) \sin \left(2 F\left(t,t_{\text{in}}\right)+\Delta _{\text{out}}\right)-\Delta \rho _{\text{out}}^2 \sin (2 g(t)) \sin \left(2 F\left(t,t_{\text{in}}\right)\right)\right)\right)\\
	& +\cos (\varphi )\sin (\theta )\left(2\rho _{in_1}\rho _{i n_2}\Delta \rho _{\text{out}}^2\left(\cos (2 g(t)) \sin \left(\Delta _{\text{in}}\right) \cos \left(2 F\left(t,t_{\text{in}}\right)\right)+2\rho _{out_1}\rho _{out_2}\left(\sin (2 g(t)) \sin \left(\Delta _{\text{in}}\right) \cos \left(2 F\left(t,t_{\text{in}}\right)+\Delta _{\text{out}}\right)\right.\right.\right.\\
	& +\left.\left.\left.\cos (2 g(t)) \cos \left(\Delta _{\text{in}}\right) \sin \left(\Delta _{\text{out}}\right)\right)\right)-\Delta \rho _{\text{in}}^2 \left(2\rho _{out_1} \rho _{out_2} \sin (2 g(t))  \sin \left(2 F\left(t,t_{\text{in}}\right)+\Delta _{\text{out}}\right)+\Delta \rho_{\text{out}}^2 \cos (2 g(t)) \sin \left(2 F\left(t,t_{\text{in}}\right)\right)\right)\right).
	\end{split}
	\end{equation}
	The denominator is the same as computed for the weak measurements of $\sigma^z$.
\end{widetext}

%
%
%
%


	\bibliography{vEinsteinDirac.bib}

\end{document}